\documentclass[review,12pt]{elsarticle}

\usepackage{amssymb}
\usepackage{amsthm}
\usepackage{amsmath}

\usepackage{mathrsfs}
\usepackage{graphicx}
\usepackage{epstopdf}
\usepackage{float}
\usepackage{caption}
\usepackage{subcaption}
\usepackage{bm}
\usepackage{bbm}
\usepackage{mathrsfs}
\usepackage{hyperref} 
\usepackage{cleveref}
\usepackage{soul}
\usepackage{accents}
\usepackage{graphicx}
\usepackage{xcolor}
\usepackage{courier} 
\usepackage{listings} 
\usepackage{tabu} 
\usepackage{longtable}
\usepackage{changepage} 
\usepackage[margin=2cm]{geometry}
\usepackage[version=4]{mhchem} 
\usepackage{lineno}
\biboptions{sort&compress} 
\usepackage{upgreek} 
\usepackage{booktabs} 

\journal{Corrosion Science}

\makeatletter
\def\@author#1{\g@addto@macro\elsauthors{\normalsize%
    \def\baselinestretch{1}%
    \upshape\authorsep#1\unskip\textsuperscript{%
      \ifx\@fnmark\@empty\else\unskip\sep\@fnmark\let\sep=,\fi
      \ifx\@corref\@empty\else\unskip\sep\@corref\let\sep=,\fi
      }%
    \def\authorsep{\unskip,\space}%
    \global\let\@fnmark\@empty
    \global\let\@corref\@empty  
    \global\let\sep\@empty}%
    \@eadauthor={#1}
}
\makeatother

\begin{document}

\begin{frontmatter}



\title{Hydrogen uptake kinetics of cathodic polarized metals in aqueous electrolytes}


\author[IC]{Livia Cupertino-Malheiros}

\author[LaSIE,TUM]{Malo Duportal}

\author[OX]{Tim Hageman}

\author[OX]{Alfredo Zafra}

\author[OX]{Emilio Mart\'{\i}nez-Pa\~neda\corref{cor1}}
\ead{emilio.martinez-paneda@eng.ox.ac.uk}

\address[IC]{Department of Civil and Environmental Engineering, Imperial College London, London SW7 2AZ, UK}

\address[LaSIE]{LaSIE, La Rochelle Universit\'e, UMR CNRS 7356, 17042 La Rochelle, France}

\address[TUM]{Department of Physics, Technical University Munich,  85748 Garching, Germany}

\address[OX]{Department of Engineering Science, University of Oxford, Oxford OX1 3PJ, UK}

\cortext[cor1]{Corresponding author.}

\begin{abstract}
We use a unique combination of electrochemical techniques to elucidate the dependency of hydrogen evolution reaction (HER) and absorption on pH and overpotential for iron and nickel. Impedance spectroscopy shows the dominance of the Volmer-Heyrovsky reaction pathway, challenging the common consideration of Volmer-Tafel dominance. Polarization slopes agree with the Volmer or Heyrovsky rate-determining step, with limitations at high overpotential. The evolution of steady-state permeation current density with overpotential is rationalised through newly-developed theory. Surface activity and absorption trends are captured. Combined with modelling, this work provides a path for quantifying hydrogen uptake and establishing an equivalent fugacity for aqueous electrolytes.\\ 
\end{abstract}


\end{frontmatter}


\section{Introduction}
\label{Introduction}

A wide range of commonly used metals and alloys are susceptible to hydrogen embrittlement (HE), which currently limits the design of structures in multiple industrial sectors such as energy and transportation \cite{Gangloff2012,Briottet2019,Djukic2019}. As any HE phenomenon is strongly related to the local hydrogen concentration \cite{Gerberich1996, Thomas2003, Kehler2007, Martinez-Paneda2016, Cupertino-Malheiros2022}, predictive models require appropriate consideration of hydrogen uptake kinetics on metal surfaces. In aqueous electrolytes, hydrogen absorption into alloys is governed by the reactions resulting in the production and desorption of the intermediate hydrogen adsorbed state \cite{Bockris1965}. The kinetics of those, and hence the absorbed lattice hydrogen concentration, are strongly dependent on the applied potential, the electrolyte and the metal surface \cite{Kehler2007, Liu2014, Martinez-Paneda2016, Lasia2019, Kim2021, Goyal2021, Karimi2023}. There is a pressing need to quantify hydrogen absorption under relevant conditions to predict hydrogen embrittlement, and to design engineering strategies and materials against it. Moreover, being able to quantify hydrogen ingress in aqueous electrolytes would allow establishing an equivalence with gaseous hydrogen uptake, enabling the use of safe and inexpensive experiments to provide the understanding needed to use hydrogen as an energy vector and leveraging decades of electrochemical hydrogen research.

A model has been recently developed that is capable, for the first time, of fully resolving the physics of hydrogen uptake by explicitly linking the electrolyte (ion transport and potential distribution) with the hydrogen ingress into metals - see Ref. \cite{Hageman2022}. Unlike previous models, this framework has the advantage of not requiring \emph{a priori} knowledge of hydrogen concentration at metal surface \cite{Cupertino-Malheiros2022, Nagao2018, Martinez-Paneda2018, Wang2006, Martinez-Paneda2020b, Yokobori2002}, nor local variations of pH and potential at surface defects, such as pits and cracks \cite{Turnbull1996, Kehler2007, Martinez-Paneda2020}. Instead, it considers the competition between all the electrochemical reactions intrinsic to the Hydrogen Evolution Reaction (HER) and the absorption-adsorption kinetics as a function of the bulk pH and potential, which are usually known. Such a model can deliver predictions of equivalent fugacity and hydrogen uptake as a function of the material, loading conditions and environment. However, its reliability relies on incorporating the appropriate input values for the rate constants and symmetry coefficients of all the competing chemical reactions, an endeavour known for its challenging nature \cite{Turnbull2015,Cooper2007,Jiang2022}.

The characterization of the hydrogen production kinetics in aqueous electrolytes has attracted great interest in the fields of HE and electrocatalysis in the past decades. Studies focused on hydrogen uptake into metals have often analyzed the dependencies of the HER-associated cathodic current and/or steady-state permeation current densities on the applied potential. This has allowed, based on some considerations, to obtain HER rate constants needed to estimate adsorbed hydrogen coverage ($\theta_\text{ads}$) \cite{Bockris1965, Iyer1988, Iyer1989,Turnbull1996, Elhamid2000, Elhamid2000b, ElAlami2006, Kehler2007, Turnbull2015, Vecchi2018, Martinez-Paneda2020}, as well as expressions of hydrogen fugacity ($f_{\text{H}_2}$), depending on the dominant HER  mechanisms on the cathodically charged metal surfaces \cite{Bockris1971, Liu2014, Venezuela2018, Liu2018, Koren2023}. Both $\theta_\text{ads}$ and $f_{\text{H}_2}$ are linked to the sub-surface lattice hydrogen concentration, the former through the relationship between the absorption and desorption rate constants, the latter through Sievert’s law. All these works have considered steady state and negligible inverse HER reactions. For further simplification, some authors have considered reactions controlled only by charge transfer \cite{Bockris1965, Turnbull1996,Martinez-Paneda2020, Iyer1988, Iyer1989,Elhamid2000,Elhamid2000b, ElAlami2006,Turnbull2015}, a reaction rate of absorption much smaller than the other steps \cite{Elhamid2000}, desorption as a general reaction disregarding its mechanism (chemical and/or electrochemical) \cite{Vecchi2018}, and, very frequently, a dominance of the Volmer-Tafel mechanism, neglecting the relative contribution of the Heyrovsky step \cite{Kehler2007, Bockris1965, Turnbull1996, Iyer1988, Iyer1989, Elhamid2000b}. 

While neglecting hydrogen absorption into metals, recent works on hydrogen electrolytic production have provided more in-depth descriptions of the HER kinetics and mechanisms on metal electrodes \cite{Kucernak2016, Watzele2018, Lasia2019, Goyal2021, Goyal2021b, Monteiro2021, Taji2022}. Experimental procedures for the determination of kinetics constants by steady state polarization, electrochemical impedance spectroscopy (EIS), and relaxation experiments have been discussed \cite{Kucernak2016, Lasia2019}. The complex dependence of electrode activities on pH and potential has been investigated \cite{Kucernak2016, Watzele2018, Goyal2021}, whereas new insights into the effect of cations and mass transport have been revealed \cite{Lasia2019, Monteiro2021, Goyal2021, Goyal2021b, Taji2022}. These studies have demonstrated the complexity of properly describing HER kinetics, despite it being one of the simplest and most frequently studied electrocatalytic reactions. As these works go beyond what has previously been considered by numerous HE investigations, they provide an exciting opportunity to better characterize HER kinetics, opening an avenue for improving predictions of hydrogen absorption and embrittlement, and for establishing equivalences with hydrogen-containing gaseous environments.

This work aims to bring recent methodological developments in the electrocatalysis community to the area of HE so as to gain new fundamental insight into hydrogen absorption from aqueous electrolytes. Experiments are conducted on Fe and Ni, two commonly embrittled metals that serve as model materials. First, their HER surface activity is investigated using steady-state cathodic polarization and EIS techniques covering a wide range of applied potentials in relatively mild electrolytes (pH 1-13). This is followed by investigations of hydrogen production and absorption by analyzing successive electrochemical permeation transients. Therefore, the current experimental methodology advances on previous works, in which only electrochemical permeation was performed \cite{Bockris1965, Iyer1989, Elhamid2000b, Liu2014, Vecchi2018}, by first providing detailed HER analyses prior to permeation tests under the same conditions (electrolytes and potentials). The effect of pH on HER and hydrogen absorption is assessed in all the experiments, using the reversible hydrogen electrode potential. EIS measurements show the dominance of the Volmer-Heyrovsky reaction pathway over Volmer-Tafel, contrary to the established wisdom. Cathodic polarization slopes are in line with the rate-determining step of Volmer (low coverage) or Heyrovsky (high coverage), with limitations due to mass transport and bubbling at high overpotentials. The steady-state permeation current density evolution with the overpotential shows a decrease in slope at $|\eta| \approx$ 0.35 V, which can be explained by new theoretical relationships that are developed from the single-step reaction rate equations, yielding expressions that depend only on the symmetry coefficients for low coverage. The results capture the similar surface activity of Fe and Ni, linked to their comparable work function, and the more favourable absorption into Ni due to its smaller heat of solution. The novel combination of electrochemical techniques used in this work has, thus, led to new insights into the HER dominant mechanism and the dependence of HER kinetics on the electrolyte pH and potential, which paves the way for the quantification of hydrogen uptake and equivalent fugacities through the use of recently developed models \cite{Hageman2022,Hageman2023}. Existing limitations are also discussed.

\section{Experimental Methods}

\label{Sec:Experimental}
\subsection{Material}

This study was performed with samples of annealed Armco\textsuperscript{\textregistered} iron and 99.9 wt.\% pure nickel. Iron was purchased as a 1 mm thick annealed sheet and nickel as a 20 mm thick cold rolled material that was then annealed at 950°C for 1 hour. Electron backscatter diffraction (EBSD) analyses revealed the expected annealed microstructures, with equiaxed randomly oriented grains with average sizes of 28 $\upmu$m and 115 $\upmu$m for iron and nickel, respectively. 

\subsection{Electrochemical tests}
\label{Subsec:ElectrochemicalTests}

\begin{figure}
     \centering
         \includegraphics[scale=0.88]{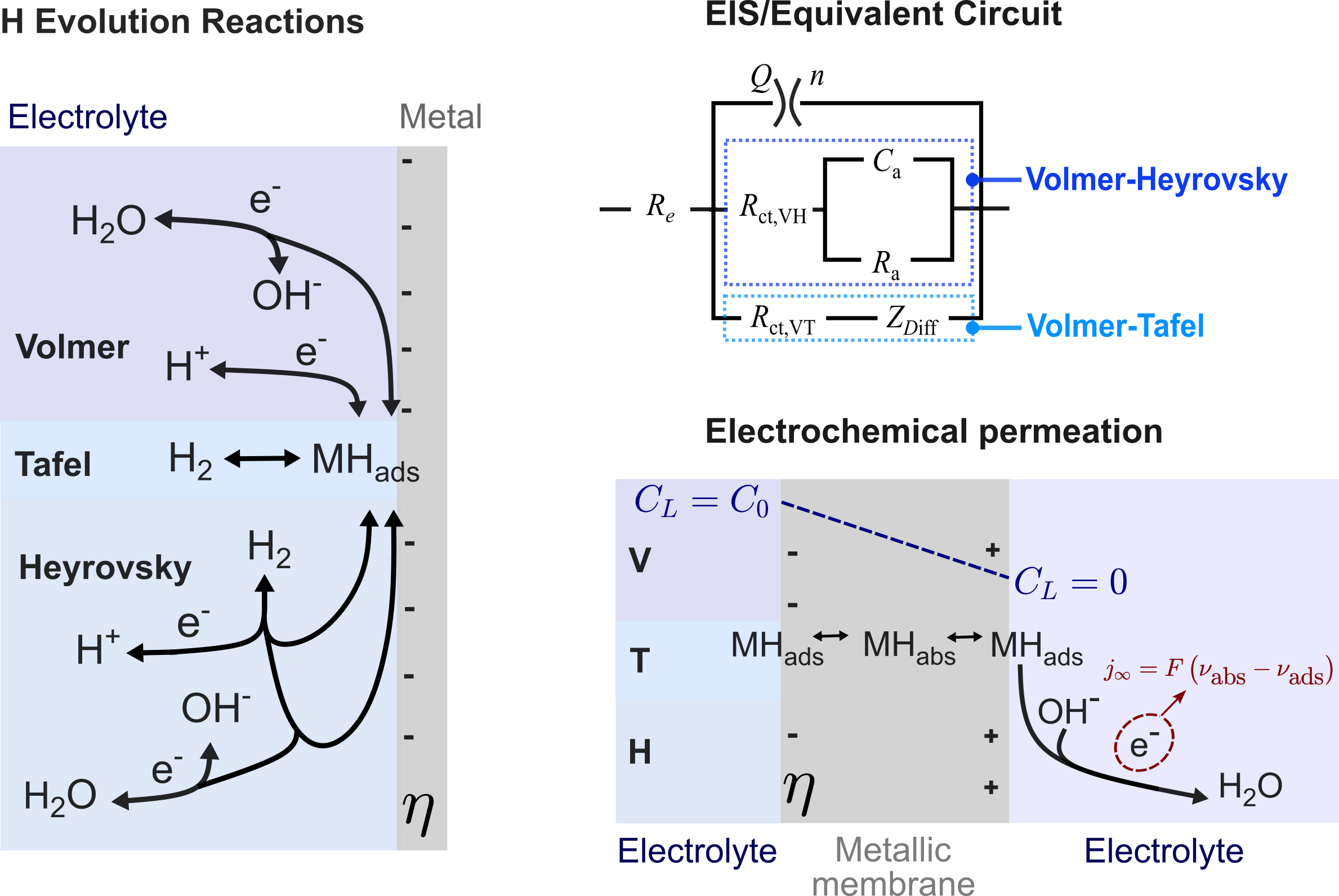}  
        \caption{Scheme of HER reactions at the metal surface (left); electrical equivalent circuit accounting for the Faradaic impedances associated with the Volmer-Heyrovsky and the Volmer-Tafel pathways, completed with an electrolyte resistor ($R_e$) and a double layer impedance (represented by $Q$ and $n$) (top-right); and hydrogen absorption and diffusion with reactions at the anodic side of the electrochemical permeation membrane (bottom-right).}
        \label{fig:scheme}
\end{figure}

Our experimental approach comprises electrochemical tests for analysing the HER mechanism and kinetics at the metal surfaces, which are then related to hydrogen absorption into metal membranes for a range of electrolyte pH and applied potential (Fig. \ref{fig:scheme}).

To this end, this work considers the usual description of HER in acid, Reaction (\ref{eq:HERa}), and alkaline Reaction (\ref{eq:HERb}) electrolytes.
\begin{equation} \label{eq:HERa}
   \ce{2H+ +2e- <=> H2}
\end{equation}
\begin{equation} \label{eq:HERb}
   \ce{2H2O +2e- <=> H2 + 2OH-}
\end{equation}

The HER mechanism includes the following steps, with their respective kinetics constants $k$: electro-adsorption (Volmer, V, Reactions (\ref{eq:Va}) and (\ref{eq:Vb})), electrochemical (Heyrovsky, H, Reactions (\ref{eq:Ha}) and (\ref{eq:Hb})) and chemical (Tafel, T, Reaction (\ref{eq:T})) desorption, and absorption (Reaction (\ref{eq:Abs})). In acid environments, the electrochemical steps are governed by proton reduction (Va and Ha), while in alkaline electrolytes they occur by water reduction (Vb and Hb). The aforementioned steps can be written as
 \begin{equation} \label{eq:Va}
   \ce{H+ +M +e- <=>C[$k$_{Va}][{$k'$_{Va}}] MH_{ads}}
\end{equation}

\begin{equation} \label{eq:Vb}
   \ce{H2O +M +e- <=>C[$k$_{Vb}][{$k'$_{Vb}}] MH_{ads} + OH-}
\end{equation}

 \begin{equation} \label{eq:Ha}
   \ce{H+ +e- + MH_{ads} <=>C[$k$_{Ha}][{$k'$_{Ha}}] M + H2}
\end{equation}

 \begin{equation} \label{eq:Hb}
   \ce{H2O +e- + MH_{ads} <=>C[$k$_{Hb}][{$k'$_{Hb}}] M + H2 + OH-}
\end{equation}

 \begin{equation} \label{eq:T}
   \ce{2MH_{ads} <=>C[$k$_{T}][{$k'$_{T}}] 2M + H2}
\end{equation}

 \begin{equation} \label{eq:Abs}
   \ce{ MH_{ads} <=>C[$k$_{A}][{$k'$_{A}}] MH_{abs}}
\end{equation}

All the electrochemical experiments were performed using Gamry 1010B potentiostats, Ag/AgCl reference electrodes, and platinum rod counter electrodes. Three aqueous solutions were employed as electrolytes: 0.05 mol/L \ce{H2SO4} (pH = 1.3), 0.5 mol/L NaCl (pH = 7.8), and 0.1 mol/L NaOH (pH = 12.8). All were made with distilled water, deaerated with nitrogen for one hour before and during the experiments. The temperature of the laboratory was set to 21 ± 2°C. The metal samples were mechanically ground and dried immediately before being immersed in the electrolyte, using SiC papers up to 4000 grit (5 $\upmu$m) finishing. 

\subsubsection{HER metal surface activity}
\label{Subsubsec:HER}

Investigation of metal surface activity was conducted through cathodic polarization (CP) and EIS in three cold-mounted Fe and Ni samples with exposed ground surfaces of 1 ± 0.1 cm\textsuperscript{2} (with the other surfaces being masked with mounting resin). 

Prior to CP testing, a constant potential was applied for one hour to reduce oxide layers that may form in alkaline environments and expose the metal surfaces.  The magnitude of these constant potentials were -0.8 and -1.1 $\text{V}_\text{Ag/AgCl}$ for Fe and -0.7 and -1.0 $\text{V}_\text{Ag/AgCl}$ for Ni in NaCl and NaOH solutions, respectively. CP curves were recorded by varying the potential at a constant rate of -0.15 mV/s from the values applied in these potentiostatic steps or from OCP for the acid electrolyte towards more negative values down to -2 $\text{V}_\text{Ag/AgCl}$. 

EIS measurements were carried out as close as possible to the protocol described by Watzele \textit{et al.} \cite{Watzele2018, Taji2022, Watzele2024}. Ranges of potential between -0.5 and -0.9 $\text{V}_\text{Ag/AgCl}$ in \ce{H2SO4}, -0.9 and -1.3 $\text{V}_\text{Ag/AgCl}$ in NaCl, and -1.1 and -1.5 $\text{V}_\text{Ag/AgCl}$ in NaOH were investigated. Samples were potentiostatically charged for 1800 seconds before the EIS tests. The frequencies analyzed ranged from 100 kHz to 10 mHz with 6 points per decade and a sinus amplitude of 10 mV. The impedance data were fitted using the Simad\textsuperscript{\textregistered} software. 

A theoretical derivation, provided by Watzele \textit{et al.} \cite{Watzele2018, Watzele2024}, is used to fit the EIS data. It considers that HER proceeds following the Volmer-Heyrovsky (VH) or the Volmer-Tafel (VT) mechanism, with these processes happening simultaneously and quasi-independently at different parts of the metal surface. From the variations in Faradaic current and coverage for a small AC-probing amplitude, the following equations for the complex impedances ($\hat{Z}$) are derived for the VH and VT mechanisms \begin{equation} \label{eq:ZVH}
  \hat{Z}_{F,VH}  = R_{ct,VH} + \frac{1}{\frac{1}{R_a} + iwC_a} 
\end{equation} 
\begin{equation} \label{eq:ZVT}
  \hat{Z}_{F,VT}  = R_{ct,VT} + \hat{Z}_{Diff}
\end{equation} where $R_{ct}$ is the charge transfer resistance, $i$ the imaginary unit, $w$ the angular frequency, $C_a$ the adsorption pseudo-capacitance, $R_a$ the adsorption resistance and $\hat{Z}_{Diff}$ the semi-infinite Warburg diffusion impedance. For the VH mechanism, Eq. (\ref{eq:ZVH}), it is important to mention that there are no real capacitances or resistances in the physical system. These elements allow expressing the AC behavior of the Volmer-Heyrovsky kinetic equations in simpler terms \cite{Watzele2018}.

Eqs. (\ref{eq:ZVH}) and (\ref{eq:ZVT}) can also be represented by the Volmer-Heyrovsky and Volmer-Tafel equivalent electric circuits (EEC) of Fig. \ref{fig:scheme}. The EEC represented in Fig. \ref{fig:scheme} is completed by the inclusion of the electrolyte resistance ($R_e$), in series with the rest of the circuit, and the impedance of the Constant Phase Element (CPE) representing the double layer behaviour ($Z_{DL}$), in parallel to the Faradaic impedances. The impedance of the CPE is given by \begin{equation} \label{eq:ZDL}
  Z_{CPE}  =  Z_{DL} = \frac{1}{Q(iw)^n}
\end{equation} with Q $\left (\text{F} \text{cm}^{-2}\text{s}^{(n -1)} \text{, F standing for farad}\right)$, attributed to the CPE, being intrinsically related to the corrector exponent \textit{n}, which expresses the deviation from a purely capacitive behaviour ($n = 0$ for purely resistive, while $n = 1$ for pure capacitance).  

\subsubsection{Hydrogen absorption}
\label{Subsubsec:absorption}

Evaluation of hydrogen uptake was performed by analysing electrochemical permeation data acquired using a Devanathan-Stachurski two-compartment electrolytic cell. The exposed area of the tested permeation membranes was 2.01 cm\textsuperscript{2} (1.6 cm diameter holder opening) and the membrane thicknesses were 0.950 ± 0.050 mm and 0.225 ± 0.015 mm for Fe and Ni, respectively. Following grinding of both surfaces, the side of the membranes exposed to the anodic cell was electroplated with a thin layer of Pd (approximately 50-nm thick) and maintained at a constant anodic potential of 0.15 $\text{V}_\text{Ag/AgCl}$ in a 0.1 mol/L NaOH (pH = 12.8) solution. Electroplating was
completed by applying a current density of -2 mA/cm$^2$ for 10 min in a solution containing 2 g/L of Pd. No changes were observed on the surfaces of the detection side of the membranes during the permeation tests. After the anodic current was stable and below 50 nA/cm\textsuperscript{2}, the entry side of the membranes was potentiostatically polarized to provide successive permeation transients. To this end, the applied potential at the entry side was changed towards more negative values when each transient reached its steady-state anodic current density. The applied potential values were between -0.5 and -1.6 $\text{V}_\text{Ag/AgCl}$ in \ce{H2SO4}, -0.9 and -1.6 $\text{V}_\text{Ag/AgCl}$ in NaCl, and -1.1 and -2.1 $\text{V}_\text{Ag/AgCl}$ in NaOH.

The permeation tests began with the membrane entry surfaces in the as-ground state. The abnormal transient phenomena reported in Ref. \cite{Liu2014} were not observed in this work. To confirm that a pre-cathodic polarisation step was not necessary, a permeation test was performed with the entry side potentiostatically polarised at -1.5 $\text{V}_\text{Ag/AgCl}$ for 48 hours in the 0.1 mol/L NaOH solution. There was no evolution of the cathodic current density at the entry side nor of the steady-state permeation current over time, proving that no surface evolution due to oxide/hydroxide reduction took place.

\section{Results}
\label{Sec:Results}
\subsection{Cathodic polarization}
\label{Subsec:Pola}

\begin{figure}[htp]
     \centering
    \begin{subfigure}{0.5\textwidth}
         \centering
         \includegraphics[scale=0.62]{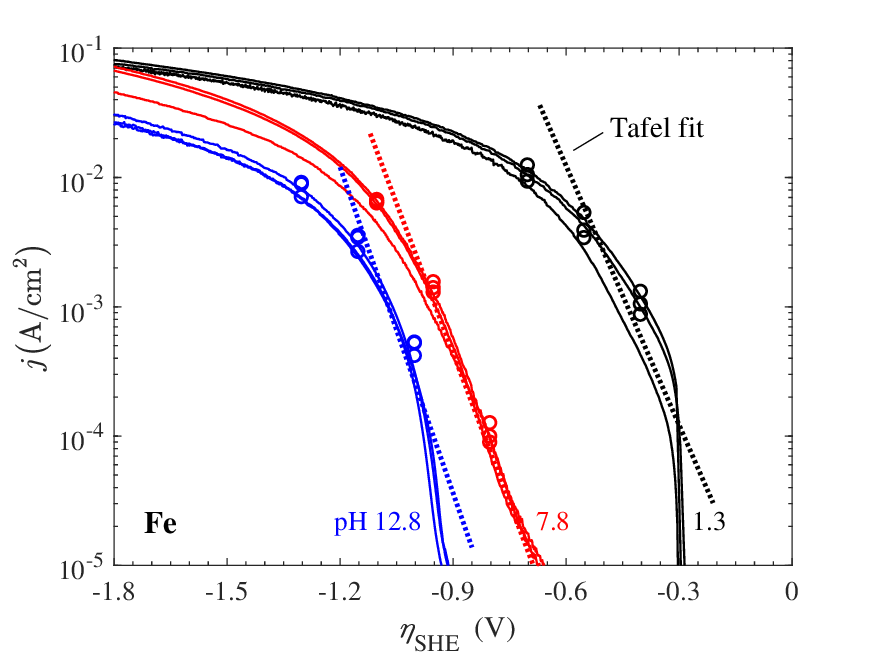}
            \caption{}
         \end{subfigure}\hfill
    \begin{subfigure}{0.5\textwidth}
         \centering
         \includegraphics[scale=0.62]{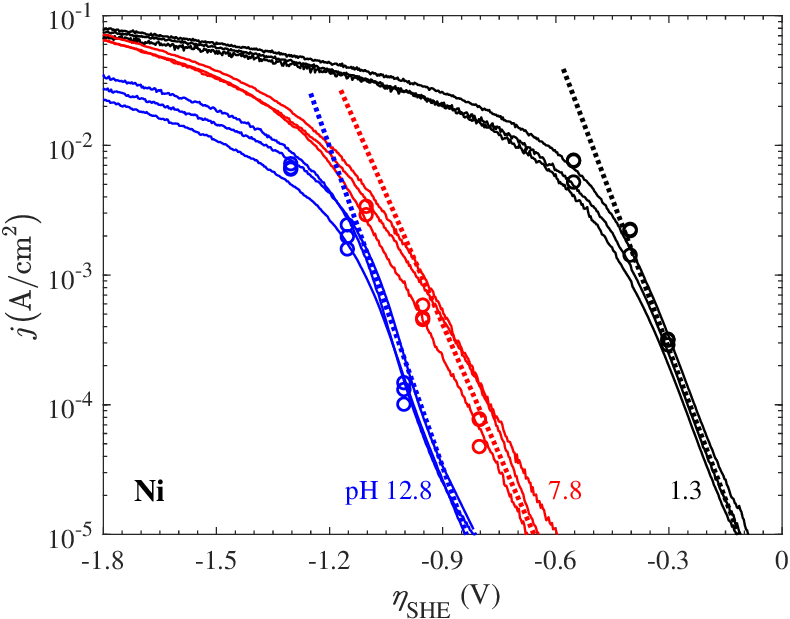}
            \caption{}
         \end{subfigure}
     \begin{subfigure}{0.5\textwidth}
         \centering
         \includegraphics[width=\textwidth]{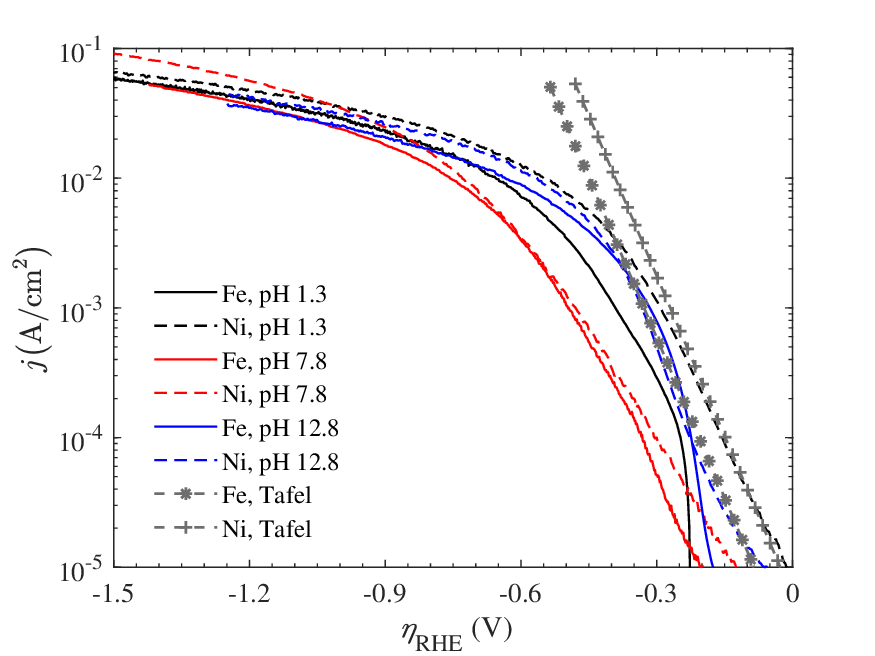}
            \caption{}
         \end{subfigure}

        \caption{Evolution of the HER-associated cathodic current density of annealed (a) Fe and (b) Ni with applied potential, and (c) Fe and Ni with applied overpotential relative to the RHE in the \ce{H2SO4} pH 1.3, NaCl pH 7.8 and NaOH pH 12.8 electrolytes. Dotted lines represent the Tafel relationships, whereas circles show the potential-current values at the end of the 1800-second potentiostatic measurements prior to each EIS test.}
    
        \label{fig:CP}
\end{figure}

Fig. \ref{fig:CP} shows the evolution of HER-associated cathodic current densities ($j$) with applied overpotential ($\eta$) for iron (Fig. \ref{fig:CP}a) and nickel (Fig. \ref{fig:CP}b) in the three electrolytes investigated. The overpotential was calculated by the difference between the applied potential ($E_\text{Ag/AgCl}$) in relation to the standard hydrogen electrode SHE ($E_\text{Ag/AgCl}+ E^0_\text{Ag/AgCl} \: \text{ where} \:\: E^0_\text{Ag/AgCl} = 0.197 \: \text{V}_\text{SHE} $) and the equilibrium potential ($E_{eq}$) \begin{equation} \label{eq:overE}
  \eta = E_\text{Ag/AgCl} + E^0_\text{Ag/AgCl} - E_{eq} 
\end{equation} 

Fig. \ref{fig:CP}a and Fig. \ref{fig:CP}b consider $E_{eq}$ equal to 0 V$_\text{SHE}$ and display the corresponding linear Tafel relationships, which express the overpotential as a function of current density according to \begin{equation} \label{eq:Tafel}
   \eta = a - b \log(j)
\end{equation} 

Alternatively, we can consider the change of potential of the SHE with pH at room temperature, i.e., $E_{eq} = RT/F \: \text{ln}([\text{H}^+]) = -0.0591$pH V$_\text{SHE}$, where $F$ is Faraday's constant, $R$ the universal gas constant and $T$ the absolute temperature. The polarization curves of both metals in relation to this reversible hydrogen electrode (RHE) are shown in Fig. \ref{fig:CP}c. In this way, the curves of Fe and Ni in the three tested electrolytes lie closer together. 

Considering the Butler-Volmer equation
\begin{equation} \label{eq:Butler}
   j = j_0 \left \{\exp{\left (-\frac{\alpha F \eta}{RT} \right)} - \exp{\left [\frac{(1-\alpha) F \eta}{RT} \right]} \right \}
\end{equation} 
and noting that the anodic process is negligible for the relatively large negative overpotentials of this work, Eq. (\ref{eq:Tafel}) can be rewritten as \cite{Bard2000} \begin{equation} \label{eq:BVTafel}
   \eta = E - E_0 = \frac{RT}{\alpha F} \ln(j_0) - \frac{RT}{\alpha F} \ln(j)
\end{equation}  where $\alpha$ is the transfer coefficient and $j_0$ the exchange current density. The slopes and intercepts of the linear Tafel relationships displayed in Fig. \ref{fig:CP}a and Fig. \ref{fig:CP}b provide average values of $\alpha$ between 0.39 and 0.49 and of $j_0$ between 10$^{-12}$ A/cm$^2$ for pH 12.8 and 10$^{-6}$ A/cm$^2$ for pH 1.3, as listed in Table \ref{CPTafel}. While $b$ and consequently $\alpha$ are related to the HER mechanism, the exchange current density provides information on the electrocatalytic activity of the metal surface. The similar cathodic polarisation behaviours of Fe and Ni are in agreement with their close position in the Volcano curve resulting from their similar interaction energies with atomic hydrogen, i.e. similar electrocatalytic activity \cite{Trasatti1972}. Considering the selected values of $j_0$ from the Trasatti review \cite{Trasatti1972} (2 and 6 x 10$^{-6}$ A/cm$^2$ for Fe and Ni, respectively) and the theoretical Tafel slope of 120 mV dec$^{-1}$, the expected Tafel relationships for both metals are shown in Fig. \ref{fig:CP}c, with relatively good agreement with the present work.

\begin{table}[H]
 \centering
  \caption{Summary of the Tafel slope (b), transfer coefficient ($\alpha$) and exchange current density ($j_0$) values for the tested conditions.}
  \label{CPTafel}
\resizebox{\textwidth}{!}{
\tiny
\renewcommand{\arraystretch}{1}
\begin{tabular}[t]{c|ccc|ccc}
\toprule
&& \text{Iron} &&&\text{Nickel} \\
\text{Parameter} & \text{\ce{0.05M H2SO4}} & \text{0.5M NaCl} & \text{0.1M NaOH} & \text{\ce{0.05M H2SO4}} & \text{0.5M NaCl} & \text{0.1M NaOH} \\
\midrule
\text{$b$ (mV dec$^{-1}$)} & 150 & 129 & 118 & 120 & 148 & 122 \\
\text{$\alpha$} & 0.39 & 0.45 & 0.49 & 0.45 & 0.39 & 0.48 \\
\text{$j_{0}$ (A/cm$^2$)} & 1.2$\times10^{-6}$  & 5$\times10^{-11}$  & 1.0$\times10^{-12}$ & 1.3$\times10^{-6}$ & 4.0$\times10^{-10}$ & 1.3$\times10^{-12}$   \\
\bottomrule
\end{tabular}}
\end{table}

\subsection{Electrochemical impedance spectroscopy}
\label{Subsec:EIS}

Polarisation curves describe the evolution of the overall HER-associated cathodic current density with the applied overpotential, providing insights into the HER mechanism and the metal catalytic activity. Complementing these data, EIS is performed to further evaluate the contribution of the different mechanisms involved in the hydrogen evolution on the studied metal surfaces \cite{Watzele2018, Taji2022}.

\begin{figure}[htp]
     \centering
    \begin{subfigure}{0.5\textwidth}
         \centering
         \includegraphics[width=\textwidth]{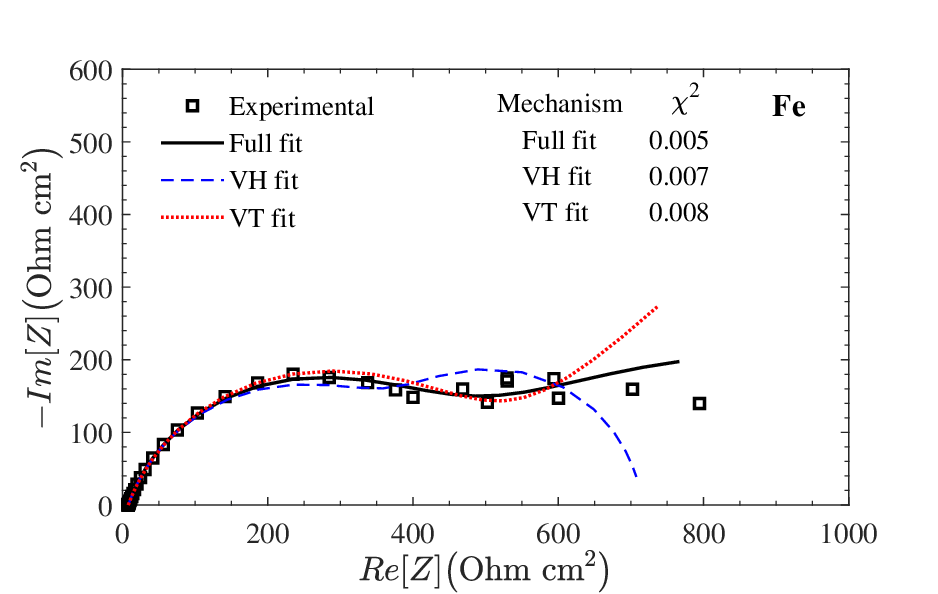}
            \caption{}
    \end{subfigure}\hfill
    \begin{subfigure}{0.5\textwidth}
         \centering
         \includegraphics[width=\textwidth]{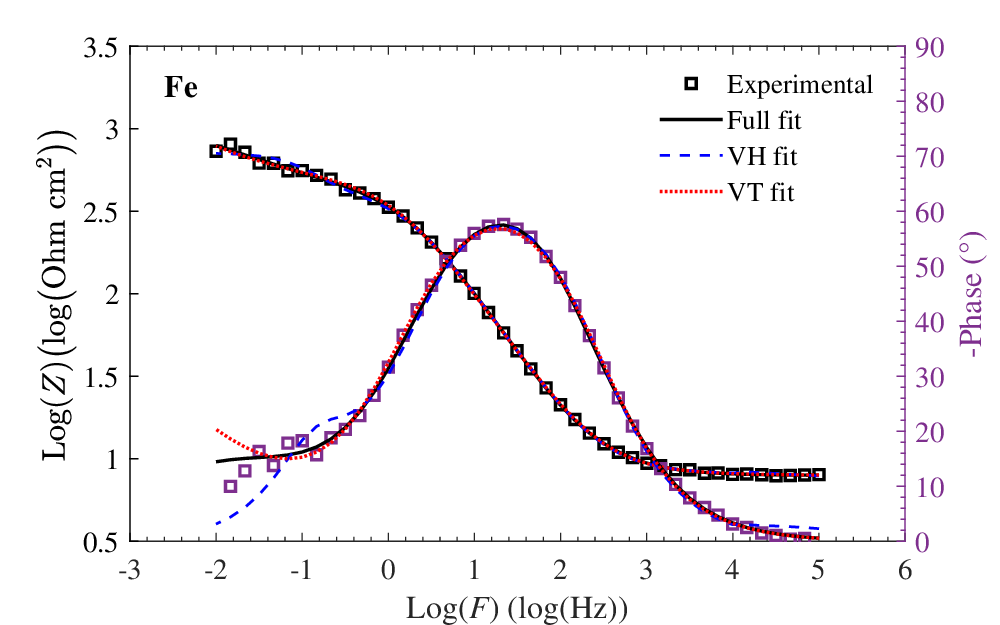}
            \caption{}
         \end{subfigure}
    \begin{subfigure}{0.5\textwidth}
         \centering
         \includegraphics[width=\textwidth]{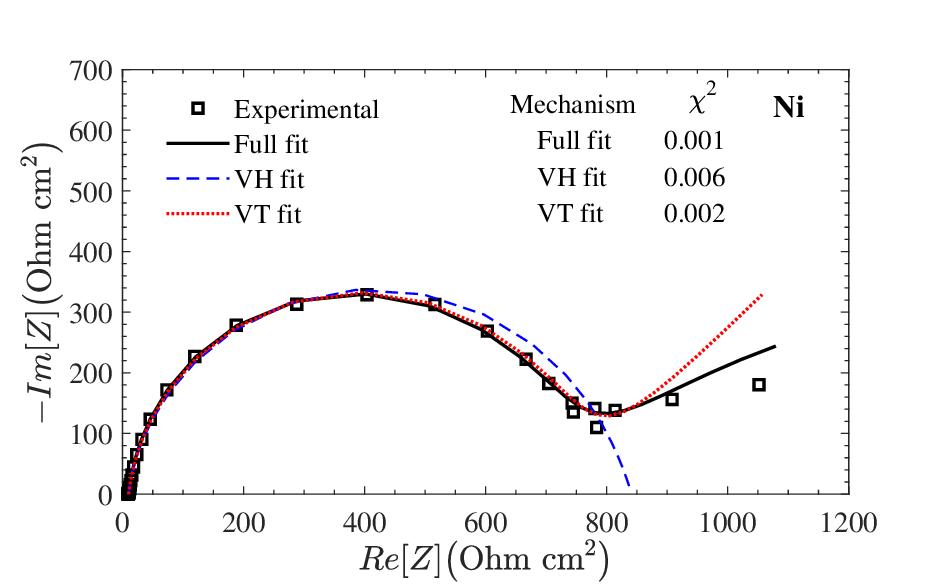}
            \caption{}
         \end{subfigure}\hfill
         \begin{subfigure}{0.5\textwidth}
         \centering
         \includegraphics[width=\textwidth]{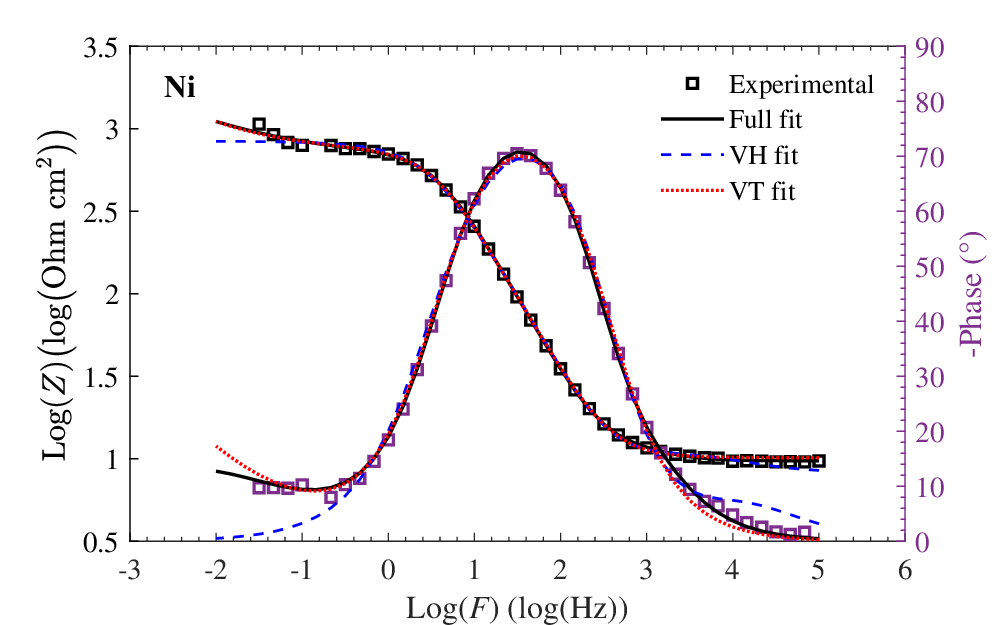}
            \caption{}
         \end{subfigure}
        \caption{Nyquist (left) and Bode (right) plots obtained in NaCl at 1.0 $\text{V}_\text{Ag/AgCl}$ for (a, b) Fe and (c, d) Ni. The dashed lines display fitting of the spectrum obtained with solely Volmer-Heyrovsky mechanism, while the dotted lines with solely Volmer-Tafel mechanism. A better fit is obtained by considering both mechanisms (solid line), as attested by the decrease of $\chi^2$.}
        \label{fig:EIS-fits}
\end{figure}

Fitting of the EIS data, acquired at the potential-current pairs represented by circles in Fig. \ref{fig:CP}a and Fig. \ref{fig:CP}b, was performed using simplified EEC attributed to solely Volmer-Heyrovsky (VH), Eq. (\ref{eq:ZVH}), and Volmer-Tafel (VT), Eq. (\ref{eq:ZVT}), with an electrolyte resistor and a double layer capacitor,  as well as the complete EEC with both VH and VT pathways. The fits obtained, and the resulting Chi-Square ($\chi^2$) values describing the quality of the fits, suggest that both mechanisms should be considered to achieve a satisfactory fit of the data, as illustrated in the Nyquist (Figs. \ref{fig:EIS-fits}a and \ref{fig:EIS-fits}c) and Bode (Figs. \ref{fig:EIS-fits}b and \ref{fig:EIS-fits}d) plots for Fe and Ni tested in the NaCl electrolyte at -1.0 $\text{V}_\text{Ag/AgCl}$. This supports the possibility of evaluating the two HER mechanisms using impedance spectroscopy and hence the complete HER equivalent electric circuit of Fig. \ref{fig:scheme} is used in the following analyses.  

It was first verified that the CPE presents a corrector exponent \textit{n} comprised between 0.8 and 1, indicating a mainly capacitive behaviour for both studied metals and all the tested conditions (applied potential and electrolyte pH). This allows, considering the double layer with a surface distribution of its time constants, a conversion of the CPE into a capacitor by using the Brug equation \cite{Brug1984, Gharbi2020} \begin{equation} \label{eq:Brug}
  C_{B}  =  Q^{1/n} ( R_e^{-1} + R_{ct}^{-1})^{(n-1)/n}
\end{equation}  
Accordingly, all the calculated capacitance values lie between 30 and 90 $\upmu$F, within the expected range for double layers.

\begin{figure}[htp]
     \centering
        
    \begin{subfigure}{0.5\textwidth}
         \centering
         \includegraphics[width=\textwidth]{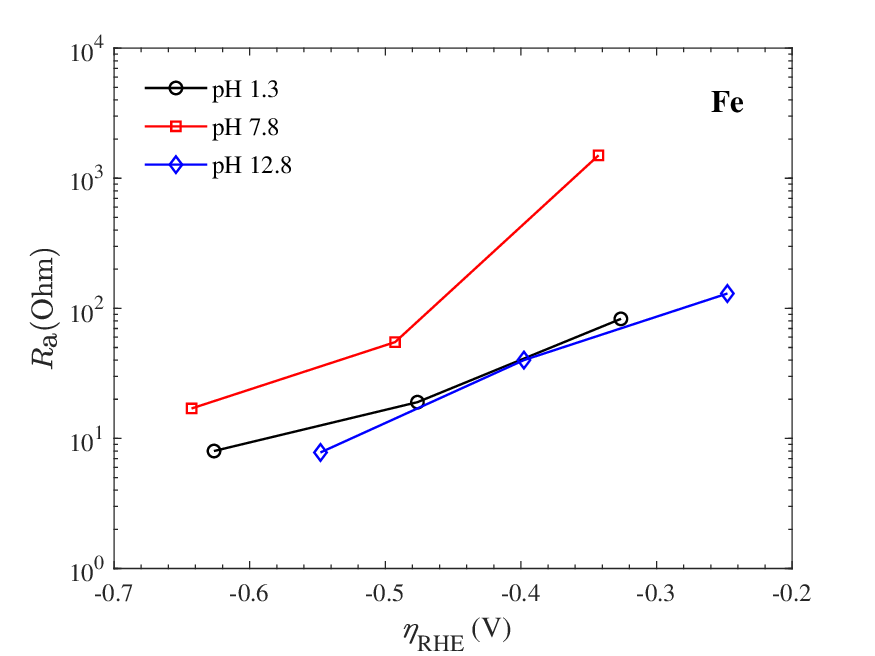}
            \caption{}
    \end{subfigure}\hfill
    \begin{subfigure}{0.5\textwidth}
         \centering
         \includegraphics[width=\textwidth]{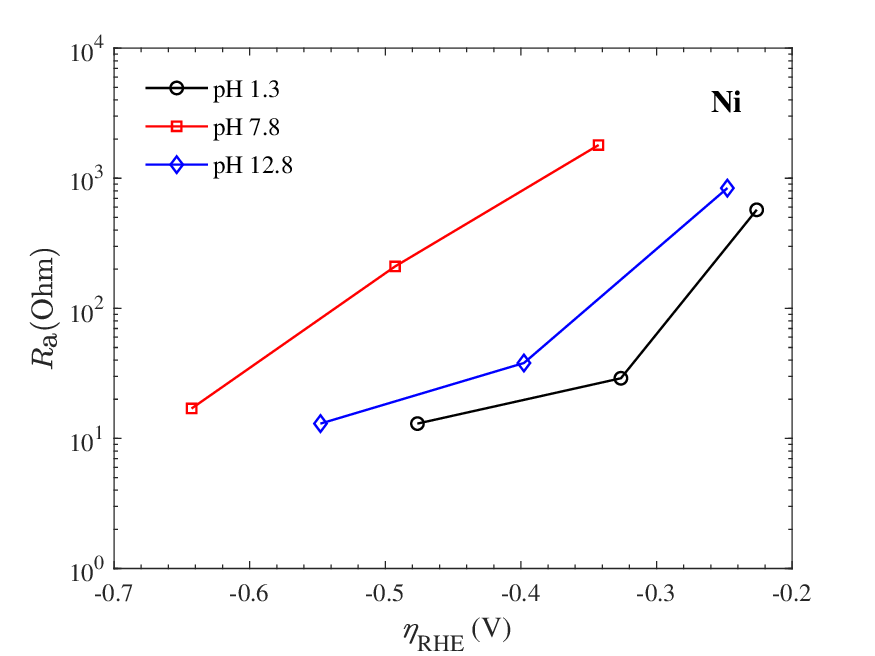}
            \caption{}
         \end{subfigure}
        \caption{Decrease of adsorption resistances with more negative overpotential values for (a) Fe and (b) Ni samples.}
        \label{fig:EIS-Ra}
\end{figure}

Among the fitted parameters from the EEC depicted in Fig. \ref{fig:scheme}, the adsorption resistance ($R_a$) is the one that presents the clearest dependency on applied overpotential. Fig. \ref{fig:EIS-Ra} shows similar tendencies, with $R_a$ decreasing as the overpotential becomes more negative, for both Fe (Fig. \ref{fig:EIS-Ra}a) and Ni (Fig. \ref{fig:EIS-Ra}b). This indicates that the adsorption of hydrogen becomes easier as the cathodic potential increases for the three tested electrolytes.

\begin{figure}[htp]
     \centering
    \begin{subfigure}{0.5\textwidth}
         \centering
         \includegraphics[width=\textwidth]{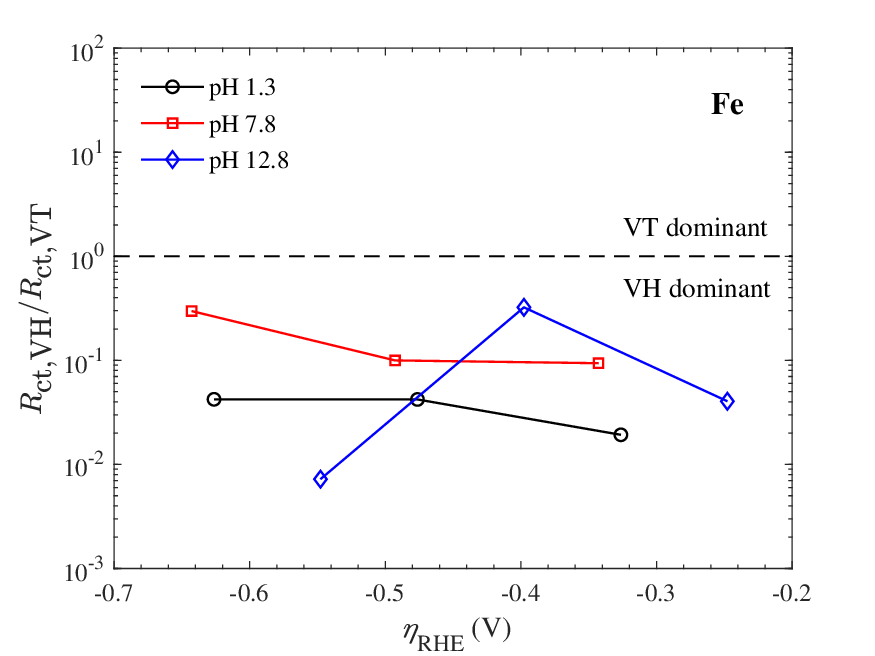}
            \caption{}
         \end{subfigure}\hfill
    \begin{subfigure}{0.5\textwidth}
         \centering
         \includegraphics[width=\textwidth]{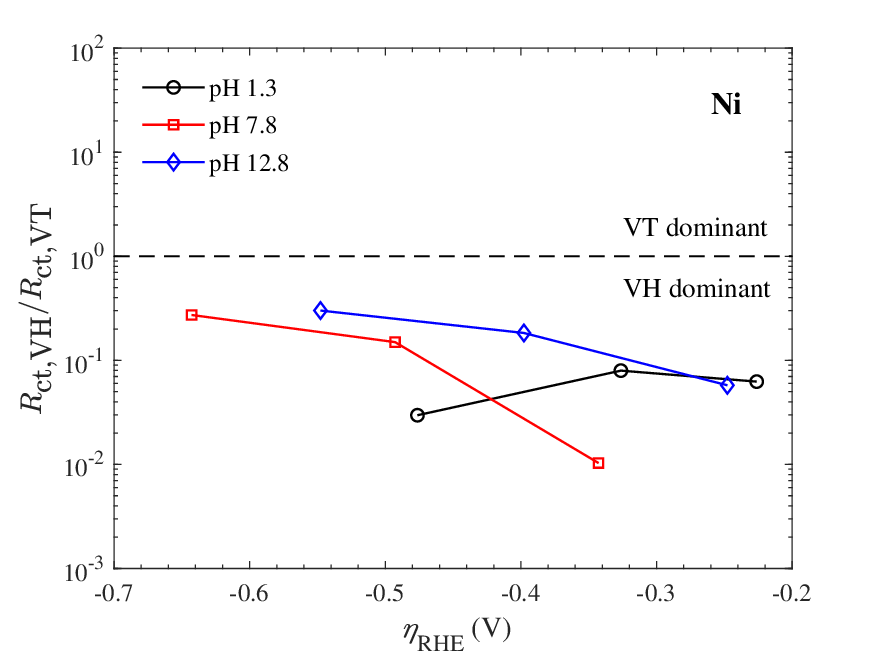}
            \caption{}
         \end{subfigure}

        \caption{Evolution of ratio of Volmer-Heyrovsky to Volmer-Tafel charge transfer resistances with the applied overpotential for (a) Fe and (b) Ni samples. }
        \label{fig:EIS-summary}
\end{figure}

The main motivation behind these EIS experiments was to provide further information on the dominant HER mechanism. The ratio between the charge transfer resistances of the VH and VT reaction pathways is used for this aim as it gives an estimate of the fraction of the current attributed to each mechanism \cite{Watzele2018}; 
\begin{equation} \label{eq:Rct0}
  R_{ct} = \frac{R_{ct,VH}}{R_{ct,VT}} = \frac{j_{VT}}{j_{VH}} 
\end{equation}
Fig. \ref{fig:EIS-summary} shows the values of this ratio as a function of applied overpotential in the three tested electrolytes for Fe (Fig. \ref{fig:EIS-summary}a) and Ni (Fig. \ref{fig:EIS-summary}b). All values lie considerably below 1, indicating that the Volmer-Heyrovsky is the dominant HER mechanism for both metals in the range of potentials investigated. This result agrees with the Tafel slopes (Fig. \ref{fig:CP}a and Fig. \ref{fig:CP}b) corresponding to the rate-determining step being either Volmer or Heyrovsky for high coverage of adsorbed hydrogen.

\subsection{Electrochemical permeation}
\label{Subsec:EP}

Successive electrochemical permeation transients were performed to investigate hydrogen absorption while avoiding trapping effects. Each permeation transient is then described by \cite{Devanathan1962, Breen1966} \begin{equation} \label{eq:Fourier}
 \frac{j_p-j_0}{j_{\infty}-j_0} = 1 + 2\sum_{n=1}^{\infty} (-1)^{n} \exp \left(-n^2\pi^2 \frac{Dt}{l^2} \right)
\end{equation} considering the sub-surface hydrogen concentration at the entry side of the membrane ($C_{0}$) constant, and given by \begin{equation} \label{eq:C0}
C_0 = \frac{j_{\infty}l}{FD} 
\end{equation}
Here, $j_p$ is the measured anodic permeation current density at a time $t$, $j_0$ is the initial current density, i.e., steady-state current density from the prior transient, $j_\infty$ is the steady-state permeation current density from the new transient, and $l$ is the thickness of the metal membrane. All the experimental permeation transients were satisfactorily fitted (R$^2\geq 0.94$) to Eq. (\ref{eq:Fourier}), using Matlab, to obtain the hydrogen diffusion coefficient $D$. The averages of the $D$ values obtained for Fe and Ni were then used to calculate $C_{0}$ for each transient.

\subsubsection{Armco\textsuperscript{\textregistered} iron}
\label{Subsec:EPFe}

\begin{figure}[htp]
     \centering
         \includegraphics[scale=0.66]{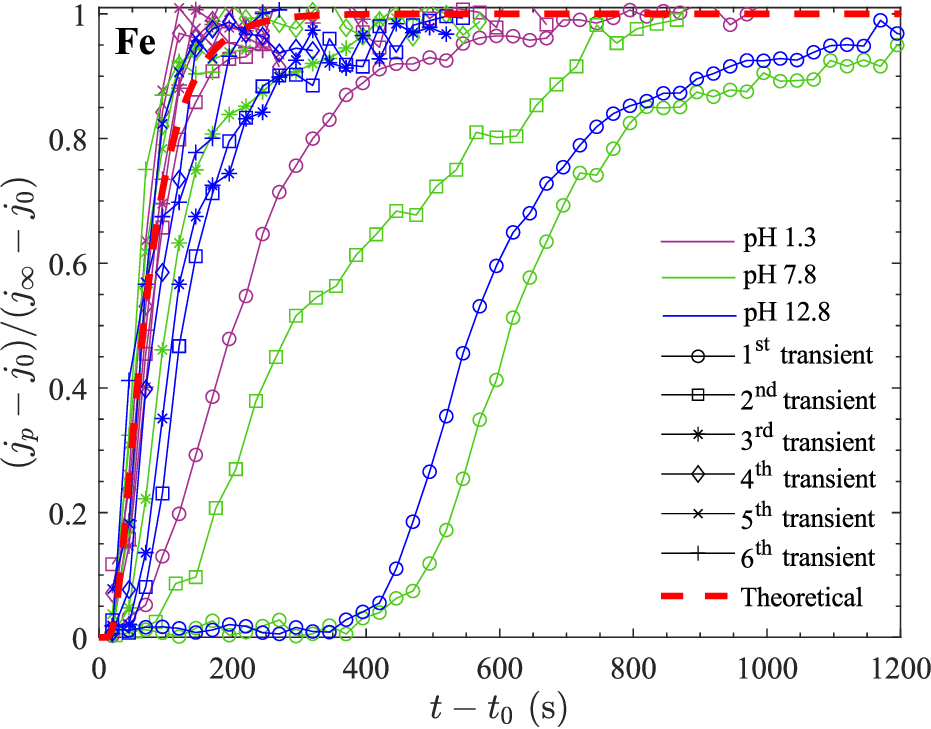}
       \caption{Representative results of normalised permeation transients of Fe for the three studied electrolytes. The theoretical transient, calculated with the average $D$ from Table \ref{EPFe}, is also displayed.}
        \label{fig:EPFe}
\end{figure}

Two permeation membranes were tested providing a total of 11 to 13 permeation transients for each electrolyte. Table \ref{EPFe} (\ref{Sec:EPdata}) lists the applied potential, cathodic and anodic current densities, diffusion coefficients, and sub-surface hydrogen concentrations for all the acquired transients. Experiments were conducted on two samples (denoted ``a'' and ``b'' in \ref{Sec:EPdata}). Fig. \ref{fig:EPFe} shows the normalised permeation transients obtained with the membrane ``a", which are representative of both samples. As expected, the first transients were more affected by trapping, exhibiting diffusion coefficients below 1$\times10^{-9}$ m$^2$/s. These values are thus not considered in the calculation of the average $D$ and the following analyses. The absorbed sub-surface hydrogen concentration at the entry side ($C_{0}$) for each transient was calculated by Eq. (\ref{eq:C0}), considering the estimated average $D$ value of 2.0 $\times 10^{-9}$ m$^2$/s, which is consistent with previous works \cite{Iyer1989, Wach1966, Choi1970, Diaz2020}. The theoretical transient displayed in Fig. \ref{fig:EPFe} was also calculated using this average diffusion coefficient. 

\begin{figure}[htp]
     \centering
         \includegraphics[scale=0.7]{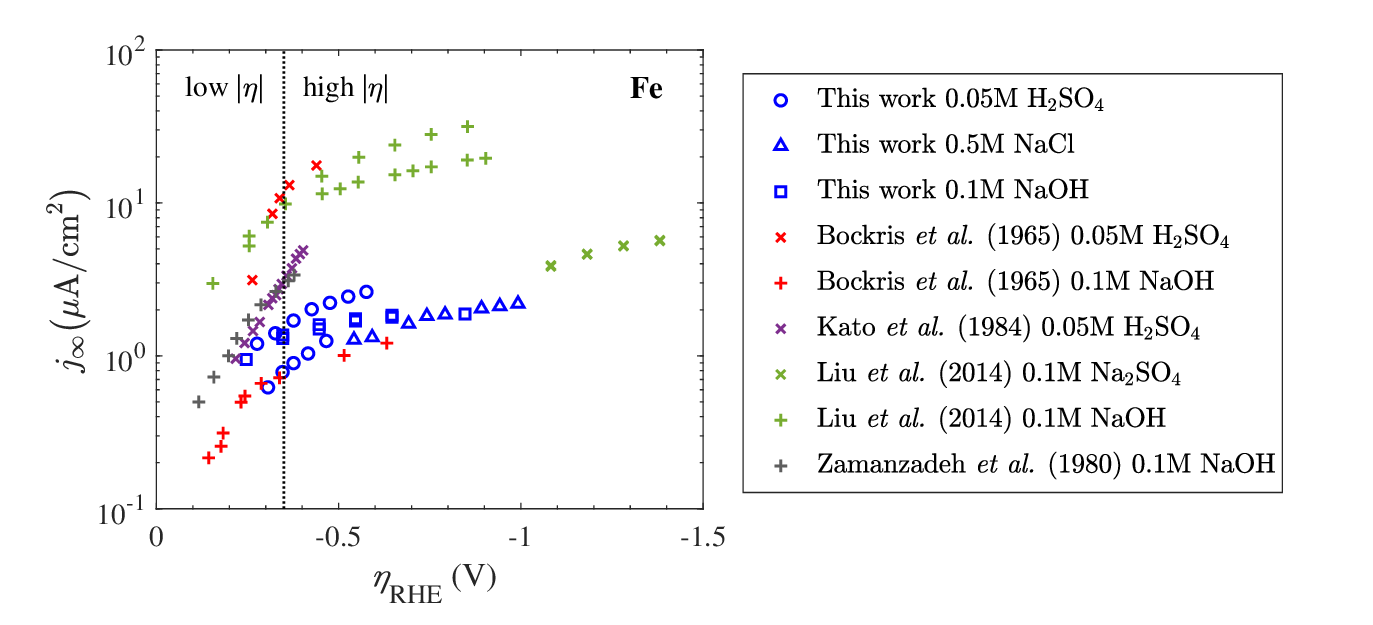}
       \caption{Evolution of the steady-state permeation anodic current density with the overpotential for Fe or low interstitial steel, as obtained from this work (blue), Bockris $et \, al$. (red) \cite{Bockris1965}, Kato $et \, al$. (purple) \cite{Kato1989}, Liu $et \, al$. (green) \cite{Liu2014}, and Zamanzadech $et \, al$. (grey) \cite{Zamanzadech1980}. The dotted line separates low and high overpotential regimes.}
        \label{fig:jmaxFe}
\end{figure}

\begin{figure}[htp]
     \centering
         \includegraphics[scale=0.7]{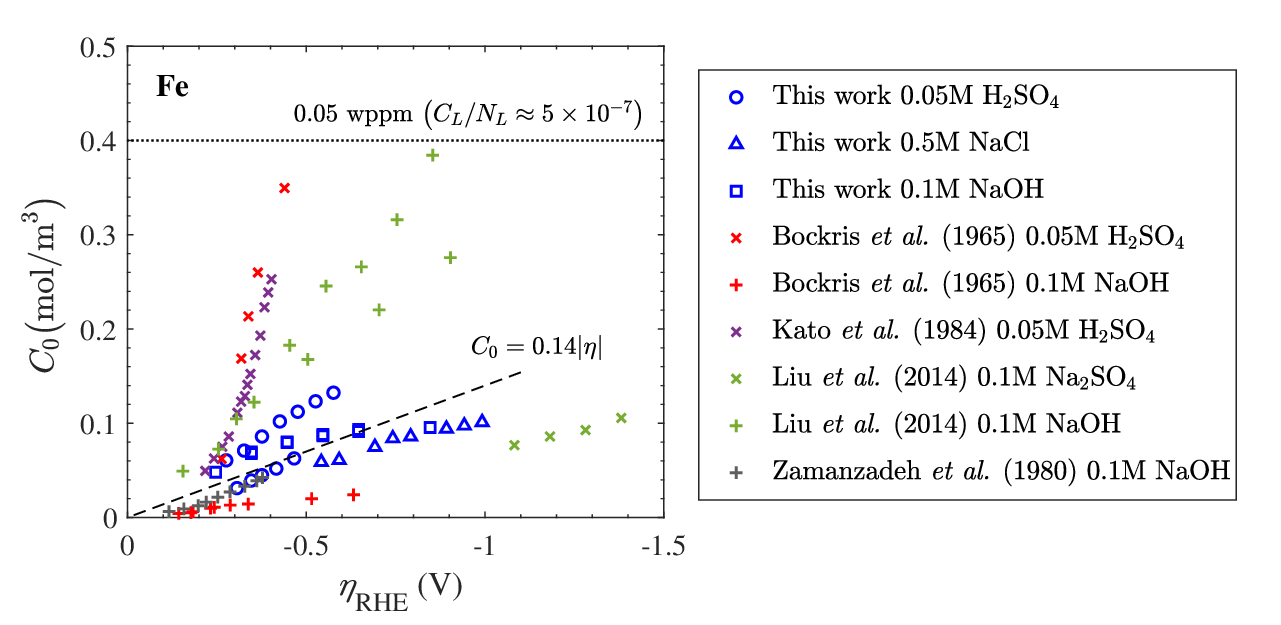}
       \caption{Evolution of the absorbed sub-surface hydrogen concentration with the overpotential for Fe or low interstitial steel, as obtained from this work (blue), Bockris $et \, al$. (red) \cite{Bockris1965}, Kato $et \, al$. (purple) \cite{Kato1989}, Liu $et \, al$. (green) \cite{Liu2014}, and Zamanzadech $et \, al$. (grey) \cite{Zamanzadech1980}.}
        \label{fig:COFe}
\end{figure}

Similar to previous works \cite{Bockris1965, Liu2014, Iyer1989, Kato1989, Zamanzadech1980}, the dependency of hydrogen absorption on the applied cathodic potential can be evaluated through the plots of the evolution of $j_\infty$ (and hence $C_{0}$) with overpotential. Fig. \ref{fig:jmaxFe} and Fig. \ref{fig:COFe} display these evolutions for the current work (blue) and previous works that used iron or low interstitial steel samples in similar electrolytes at room temperature. 

For the steady-state current density, two regimes were previously observed. They are separated in Fig. \ref{fig:jmaxFe} by a dotted line at $\eta$ = -0.35 V. For potentials less negative than this value, the ``low $|\eta|$" region in Fig. \ref{fig:jmaxFe}, there is a steeper evolution of $j_\infty$ with overpotential. The slopes $\partial{\ln(j_{\infty})} / \partial{\eta}$ are around -9 for the works by Bockris $et \, al$. \cite{Bockris1965}, Kato $et \, al$. \cite{Kato1989} and Zamanzadech $et \, al$. \cite{Zamanzadech1980}, whereas Liu $et \, al$. \cite{Liu2014} and the few points from the present work on this regime suggest slopes of $\partial{\ln(j_{\infty})} / \partial{\eta}=-6$ and $\partial{\ln(j_{\infty})} / \partial{\eta}=-4$, respectively. On the other hand, for the ``high $|\eta|$" region in Fig. \ref{fig:jmaxFe}, with potentials more negative than -0.35V, all reported measurements present a more gradual evolution of hydrogen steady-state flux with overpotential, showing slopes between $\partial{\ln(j_{\infty})} / \partial{\eta}=-2$ and $\partial{\ln(j_{\infty})} / \partial{\eta}=-1$. 

Regarding the dependency of absorbed hydrogen concentration on overpotential, the data from this work for the three tested electrolytes (blue symbols) lie around the linear relationship $C_0 = 0.14|\eta_\text{RHE}|$, displayed as a dashed line in Fig. \ref{fig:COFe}. There is scatter in the literature data with a slope between 0.04 and 0.6 if linear relationships are considered. All the $C_{0}$ acquired data, from permeation of this and previous works, are below 0.05 wppm. This corresponds to dilute solutions with occupancy up to $C_L/N_L \approx 5\times10^{-7}$, considering tetrahedral site occupancy in the iron bcc cell \cite{Jiang2004}.

\subsubsection{Pure nickel}
\label{Subsec:EPNi}

\begin{figure}[htp]
     \centering
         \includegraphics[scale=0.62]{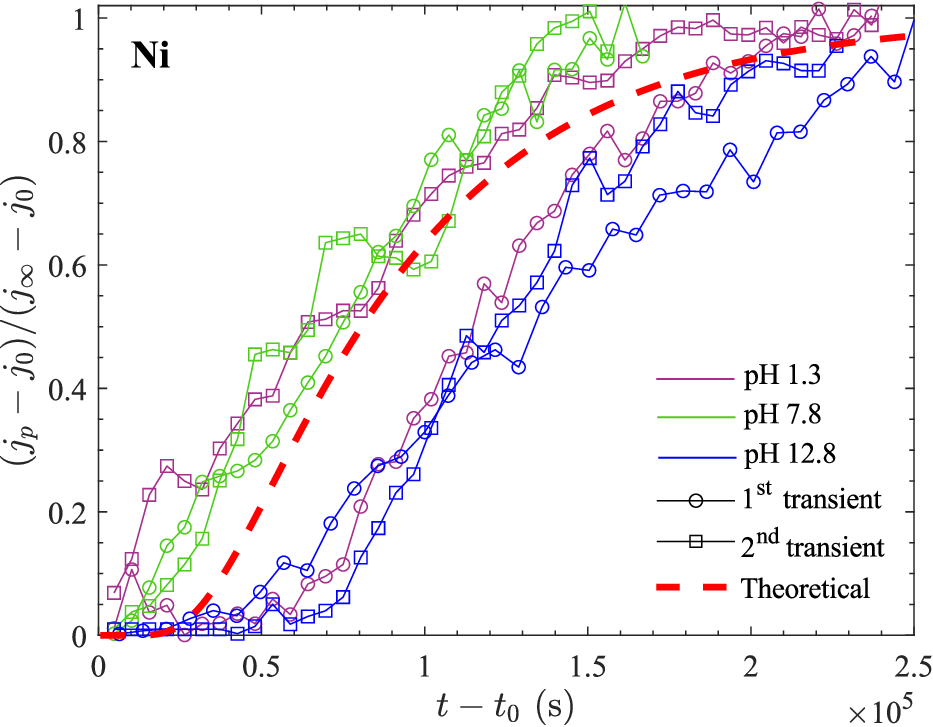}
       \caption{Representative results of normalised permeation transients of Ni for the three studied electrolytes. The theoretical transient, calculated with the average $D$ from Table \ref{EPNi}, is also displayed.}
        \label{fig:EPNi}
\end{figure}

As per the Fe case, two Ni permeation membranes were tested, ``a" and ``b", per electrolyte studied. Due to the significantly slower diffusion kinetics of hydrogen in Ni compared to Fe, only two transients were obtained for each membrane. Table \ref{EPNi} (\ref{Sec:EPdata}) provides a summary of the applied potential, cathodic and anodic current densities, diffusion coefficients, and sub-surface hydrogen concentrations for all the acquired transients. The diffusion coefficient values obtained by fitting these transients were more dispersed, with no clear increasing trend in the second transient. This dispersion may be linked to microstructural heterogeneities, especially fast diffusion and/or trapping at grain boundaries, due to the small thickness of the samples (around 0.23 mm) compared to their relatively large grain size (115 $\upmu$m) \cite{Oudriss2012}. Additionally, the very low values of the anodic current densities may also have contributed to increasing the dispersion of the permeation transients. Despite these effects, the average diffusion coefficient of 9.1 $\times 10^{-14}$ m$^2$/s is in line with previous works for Ni using thermal desorption analysis and gaseous permeation \cite{Lee1986, Kuhn1991, Robertson1973} and is used in the present work for calculating the $C_{0}$ values. Fig. \ref{fig:EPNi} shows the normalised transients for the ``a" membranes, representative of both samples, along with the theoretical transient calculated using the average diffusion coefficient 9.1 $\times 10^{-14}$ m$^2$/s. 

\begin{figure}[htp]
     \centering
         \includegraphics[scale=0.67]{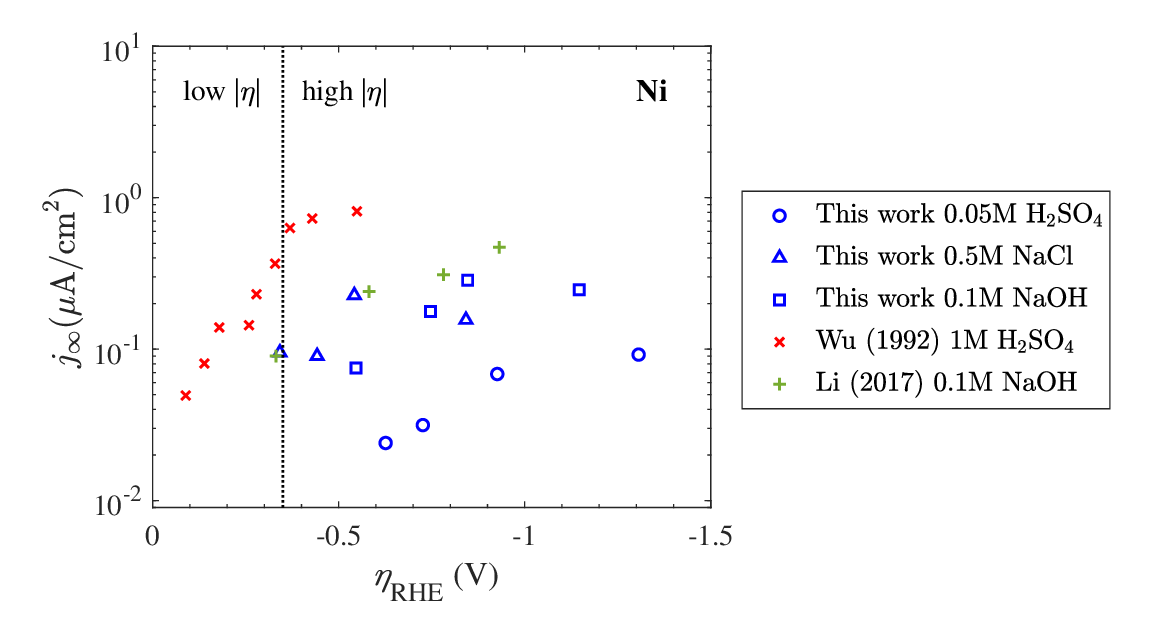}
       \caption{Evolution of the steady-state permeation anodic current density with the overpotential for Ni from this work (blue), Wu (red) \cite{Wu1992} and Li (green) \cite{Li2017}. Dotted line separates low and high overpotential regimes.}
        \label{fig:jmaxNi}
\end{figure}

\begin{figure}[htp]
     \centering
         \includegraphics[scale=0.67]{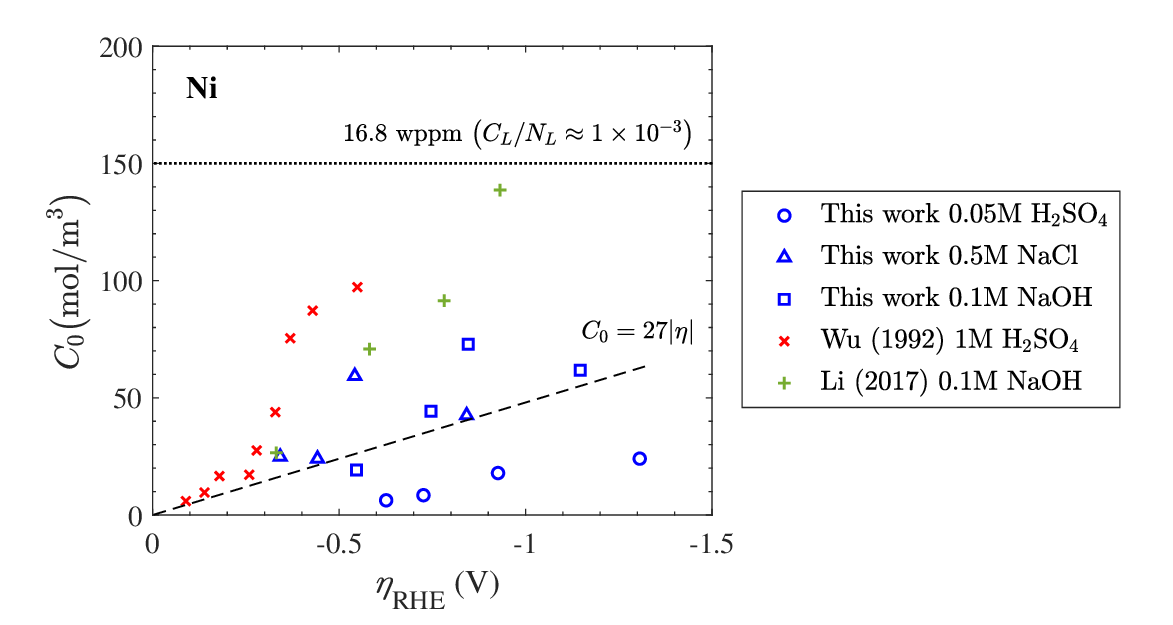}
       \caption{Evolution of the absorbed sub-surface hydrogen concentration with the overpotential for Ni from this work (blue), Wu (red) \cite{Wu1992} and Li (green) \cite{Li2017}.}
        \label{fig:CONi}
\end{figure}

Figures \ref{fig:jmaxNi} and \ref{fig:CONi} show the evolution of $j_\infty$ and $C_{0}$ with overpotential for the permeation transients of this work, along with data from two previous works \cite{Wu1992, Li2017}, for pure Ni at room temperature. Again, the data at the “high $|\eta|$” region ($|\eta| > 0.35$ V) have slopes ($\partial{\ln(j_{\infty})} / \partial{\eta}$) between -2 and -1, whereas the only data points located in the “low $|\eta|$” region from Wu's work \cite{Wu1992} have a steeper slope of -7.5. The $C_{0}$ variation with overpotential (Fig. \ref{fig:CONi}) from this work is more dispersed for Ni than Fe, lying around the linear relationship $C_0 = 27|\eta_\text{RHE}|$ (dashed line), with significantly lower values in the acid electrolyte. $C_{0}$ data from previous works increase more rapidly with overpotential at the “high $|\eta|$” region. Maximum reported values of $C_{0}$ are below 16.8 wppm. This is equivalent to an occupancy of around $C_L/N_L \approx 1\times10^{-3}$, considering lattice hydrogen occupancy in octahedral sites of the nickel fcc unit cell \cite{Wipf1997b}. 

\section{Discussion}
\label{Sec:Discussion}

\subsection{Metal surface activity and dominant HER mechanism}
\label{Subsec:HER}

The polarisation curves provide information on the dependency of the metal surface activity on pH and on the active HER mechanism. The observed decrease in current density associated with HER (and thus $j_0$) with increasing pH is in line with previous observations of slower reaction kinetics for more alkaline environments \cite{Zheng2016, Huang2021}. This influence of electrolyte pH was appropriately incorporated in the present work by considering the change of potential of the standard hydrogen electrode with pH; i.e., using the Reversible Hydrogen Electrode overpotential. 

Tafel slopes of the polarisation curves, ranging between 120-150 mV dec$^{-1}$ (Table \ref{CPTafel}), suggest that the HER rate-determining step is Volmer or even Heyrovsky with high surface coverage ($\theta_\text{ads}>$ 0.6) \cite{Shinagawa2015}. On the other hand, at larger cathodic overpotential ($|\eta_\text{RHE}| > \; \sim 0.6$ V), the current density signal becomes noisier and the curves flatten out with $b$ above 2000 mV dec$^{-1}$, which does not correspond to any theoretically expected value of the Tafel slope for HER. Limitations imposed by mass transfer and hydrogen bubbles, which reduce the exposed area of the metal surface, are probably the main reasons to explain this behaviour at large overpotential. The polarization curves thus highlight the challenge of investigating HER kinetics in cathodic potential regimes commonly used for HE studies. This indicates the need to explore more sophisticated working electrodes, such as rotating disks and microelectrodes, the benefits of which are already being commonly explored in electrocatalysis studies (see Refs. \cite{Zheng2016, Watzele2018, Huang2021}). 

These limitations at larger overpotentials, especially bubbles, restrict the potential range in which EIS measurements were performed. Even so, the measurements were performed in three potential values for each studied electrolyte and metal, comprising the entire Tafel regime (from $j$ of $10^{-4}$ to $10^{-2}$ A/cm$^2$). In Fig. \ref{fig:CP}, a very good agreement is observed between the polarization curves and the ($\eta \:$,$\: j$) pairs after 1800 seconds of potentiostatic charging. This indicates that the polarization scan rate of 0.15 mV/s was slow enough to attain a stationary state. 

Fitting the EIS data provided interesting results. Firstly, the double layer capacitance calculated from the CPE (Eq. (\ref{eq:Brug})), ranging from 30 to 90 $\upmu$F/cm$^2$, indicates that the metal surfaces are exposed, as lower capacitance values would be expected if oxide layers were present (see \cite{Pinkowski1996}). Secondly, the decrease in $R_a$ (Fig. \ref{fig:EIS-Ra}), representative of the resistance to hydrogen adsorption at the electrode surface for the Volmer-Heyrovsky mechanism, reveals a favouring of this mechanism as $|\eta|$ increases. Finally, the ratio between the charge transfer resistances shows a predominance of the Volmer-Heyrovsky over the Volmer-Tafel mechanism. As there is no clear trend in the evolution of this ratio with the overpotential in the range investigated, it was not possible to observe in this work evidence of the transition suggested in the literature \cite{Bockris1965, Liu2014} from the V-T to V-H mechanism as the applied potential becomes more negative. While this work does not claim to provide proof that this transition of mechanism  does not occur, as this would require further EIS investigation, it does challenge the considerations of dominant HER mechanisms, often based on changes in the slopes of cathodic polarization curves \cite{Tang2020}. Especially, as discussed above, considering that polarisation data at large overpotential ($|\eta_\text{RHE}| > \; \sim 0.6$ V) can be unreliable and therefore an assessment of the rate-determining step at this regime needs improvements to the methodological approach. 

The two metals studied, iron and nickel, presented very similar EIS and steady-state current-overpotential data. These results corroborate previous works \cite{Trasatti1972, Conway1957, Zeradjanin2016} showing a strong dependency of HER kinetics on the metal electronic structure, with similar reported values of work function, metal-hydrogen bond strength and exchange current density for both metals.

\subsection{Evolution of electrochemical permeation with overpotential}
\label{Subsec:Permeation}

Electrochemical permeation allows HER at metal surfaces, Reactions (\ref{eq:Va}) to (\ref{eq:T}), to be coupled  with hydrogen absorption into metals, Reaction (\ref{eq:Abs}). HER at the entry surface results in a specific coverage ($\theta_\text{ads}$) of adsorbed atoms, providing a hydrogen flux at the exit side, whose steady-state value is proportional to the concentration of hydrogen that is absorbed into the permeation membranes (Fig. \ref{fig:scheme}b). For small surface coverage ($\theta_\text{ads} \ll$ 1) and diluted solutions ($C_L \ll$ $N_L$), this can be described by \begin{equation} \label{eq:jinf2}
  j_{\infty} = F(\nu_{A}- \nu^{'}_{A}) = F(k_{A}\theta_\text{ads} - k^{'}_{A}C_0) 
\end{equation} where $\nu_{A}$ and $\nu^{'}_A$ are the reaction rates of the forward (surface to bulk) and reverse (bulk to surface) absorption reaction (\ref{eq:Abs}).

As $C_0$ is a function of $j_\infty$ (Eq. (\ref{eq:C0})), the slopes of Fig. \ref{fig:jmaxFe} and Fig. \ref{fig:jmaxNi} also represent the evolution of adsorbed hydrogen coverage ($\theta_\text{ads}$) with overpotential, as follows \cite{Bockris1965} \begin{equation} \label{eq:jinf3}
  \frac{\partial{\ln(j_{\infty})}}{\partial{\eta}} = \frac{\partial{\ln(\theta_\text{ads})}}{\partial{\eta}}
\end{equation}

By considering the reverse reactions and the absorption rate to be negligible compared to the direct Volmer, Heyrovsky and Tafel reactions, the net reaction rate ($r$) related to coverage of adsorbed hydrogen ($\theta_\text{ads}$) is given by: \begin{equation} \label{eq:r1}
    r = \frac{q}{F}\frac{d\theta_\text{ads}}{dt} = \nu_V - \nu_H - 2\nu_T
\end{equation} where $q$ is the charge required to fully form a monolayer of $\theta_\text{ads}$, and $\nu_V$, $\nu_H$ and $\nu_T$ are the reaction rates of the Volmer, Heyrovsky and Tafel steps. These are given below, with the Volmer and Heyrovsky steps being described by the acid ($\nu_{Va}$, $\nu_{Ha}$) and basic ($\nu_{Vb}$, $\nu_{Hb}$) reaction rates. 
\begin{equation} \label{eq:vVa}
    \nu_{Va} = k_{Va}C_{\text{H}^+}\exp \left(\frac{-\beta_{Va}\eta F}{RT} \right)(1-\theta_\text{ads})
\end{equation}

\begin{equation} \label{eq:vHa}
    \nu_{Ha} = k_{Ha}C_{\text{H}^+}\exp \left(\frac{-\beta_{Ha}\eta F}{RT} \right)\theta_\text{ads} 
\end{equation}

\begin{equation} \label{eq:vVb}
    \nu_{Vb} = k_{Vb}\exp \left(\frac{-\beta_{Vb}\eta F}{RT} \right)(1-\theta_{ads}) 
\end{equation}

\begin{equation} \label{eq:vHb}
    \nu_{Hb} = k_{Hb} \exp \left(\frac{-\beta_{Hb}\eta F}{RT} \right)\theta_\text{ads} 
\end{equation}

\begin{equation} \label{eq:vT}
    \nu_{T} = k_T \theta_\text{ads}^2
\end{equation}

For steady-state conditions ($\text{d}\theta_\text{ads}/\text{d}t =0$) and small surface coverage ($\theta_\text{ads} \ll$ 1), Eqs. (\ref{eq:jinf3}) to (\ref{eq:vT}) yield to expressions for the slopes of Figs. \ref{fig:jmaxFe} and \ref{fig:jmaxNi} depending only on the symmetry coefficients $\beta$, see \cref{app:B} for a full derivation. At room temperatures, these are: \begin{equation} \label{eq:jinf3b} \text{Volmer-Heyrovsky:} \quad  \frac{\partial{\ln(j_{\infty})}}{\partial{\eta}} = \frac{(\beta_H - \beta_V) F}{RT} \approx 39 \: (\beta_H - \beta_V)
\end{equation} \begin{equation} \label{eq:jinf3c} \text{Volmer-Tafel:} \quad
  \frac{\partial{\ln(j_{\infty})}}{\partial{\eta}} = \frac{-\beta_V F}{2RT} \approx -19.5 \: \beta_V
\end{equation} 

For small overpotentials, in the ``low $|\eta|$" region, the higher $\partial{\ln(j_{\infty})}/\partial{\eta}$  slopes around -9 from earlier works \cite{Bockris1965, Kato1989, Zamanzadech1980} were previously associated with the Volmer-Tafel mechanism. In this case, the polarization transfer coefficient $\alpha$ of roughly 0.5 would indicate that the reaction rate is determined by the Volmer step for low adsorbed coverage, being equivalent to $\beta_V$, resulting in a slope between -10 and -9 according to Eq. (\ref{eq:jinf3c}). On the other hand, it was previously proposed that the decrease in $\partial{\ln(j_{\infty})}/\partial{\eta}$ slopes at higher overpotential, in the ``high $|\eta|$" region, was linked to a switch to a Volmer-Heyrovsky dominant mechanism. Although these explanations are possible, the present discussion reveals other possibilities to explain these results. For low adsorbed coverage, the $\partial{\ln(j_{\infty})}/\partial{\eta}$ slopes could be explained by the differences between the symmetry coefficients $\beta_H$ and $\beta_V$, according to Eq. (\ref{eq:jinf3b}), where $\beta_V - \beta_H \approx$ 0.2 and 0.05 could explain the slopes of the low and high $|\eta|$ regions, respectively. It is likely, though, that the low coverage assumption is not valid in the entire range of overpotentials, which would result in more complex relationships between the slopes and the mechanism. And the same could be said if the reverse reactions cannot be neglected. Finally, similarly to the polarization slopes, at higher overpotentials the permeation data is likely to be significantly limited by mass transport and bubbles, which could also contribute to the observed change in the $\partial{\ln(j_{\infty})}/\partial{\eta}$ rather than by only the HER dominant pathway.

In addition to providing information on the active HER mechanism, the linear relationships between $\ln(j_{\infty})$ and $\eta$ have also been used to establish ``equivalent fugacity" relationships, in an attempt to link hydrogen absorption from electrochemical and gaseous environments \cite{Liu2014}. However, the dispersion of the slope and intercept values from this and previous works (Figs. \ref{fig:jmaxFe} and \ref{fig:jmaxNi}) together with the complexity of experimental HER evaluation, highlight the challenge of obtaining these ``equivalent fugacity" relationships reliably using Devanathan-Stachurski electrochemical permeation devices. 

The final results regard the concentration of absorbed hydrogen in iron and nickel. These metals, with similar metal-hydrogen bond strengths, show similar behavior regarding HER kinetics and hence adsorbed hydrogen coverage for a given overpotential. Their differences in hydrogen concentration (Figs. \ref{fig:COFe} and \ref{fig:CONi}) are thus mainly attributed to their solubilities, which are linked to the standard Gibbs free energy of hydrogen dissolution in the interstitial sites of their crystal structure, especially its enthalpy term (around 0.16 eV and 0.28 eV for Ni and Fe, respectively) \cite{Wipf1997, griessen1988}. This heat of solution involved during the absorption of hydrogen atoms depends not only on the metal electronic structure, but also on the distance of the hydrogen to the nearest neighbor atoms in the respective occupied interstitial sites. This distance is larger for nickel's octahedral sites than in iron's tetrahedral sites, contributing to its lower heat of solution and higher solubility \cite{griessen1988}. The hydrogen concentrations reported in this work, which are around 200 times higher in nickel than in iron, are in agreement with their solubility ratio at room temperature under equilibrium conditions \cite{Wipf1997}. Analogously, it would also be possible to attribute the difference in hydrogen solubility between these metals using their ratios of absorption ($k_{\text{A}}$) and adsorption ($k'_{\text{A}}$) kinetics constants (Reaction (\ref{eq:Abs})). However, reliable values of these constants are yet to be obtained.

\section{Conclusions}
\label{Sec:Conclusion}

By performing polarization, EIS and electrochemical permeation on Fe and Ni metal surfaces over a broad pH and overpotential range, this study has demonstrated that:

\begin{itemize}
  \item The influence of electrolyte pH on the HER kinetics and absorbed hydrogen concentration could be incorporated by using the Reversible Hydrogen Electrode potential.
  \item Tafel slopes could correspond to either Volmer (low $\theta_{ads}$) or Heyrovsky (high $\theta_\text{ads}$) determining the HER rate, whereas  EIS shows no evident change in mechanism, with a dominance of the Volmer-Heyrovsky pathway. 
  \item As the potential becomes more negative, polarization curves flatten out to slopes that do not correspond to any theoretically expected value for HER.
  \item Change in $\partial{\ln(j_{\infty})}/\partial{\eta}$ suggests shifts in the HER mechanism. Slopes of around -9 for $|\eta| < 0.35$ V can be related to the Volmer-Tafel mechanism with low coverage, but can also be explained by the Volmer-Heyrovsky path depending on the single-steps $\beta$ values. 
  \item Complex relationships between $j_{\infty}$ and $\eta$ at higher coverage and when reverse reactions cannot be disregarded, together with the limitations arising from mass transport and bubbling, can prevent the correct mechanism interpretation from permeation results. 
  \item The electrochemical experiments correctly capture the comparable metal surface activity of Fe and Ni, but with hydrogen solubility in Ni being around 200 times higher than in Fe. 
\end{itemize}

Therefore, this work challenges previous interpretations of changes in Tafel and $\partial{\ln(j_{\infty})}/\partial{\eta}$ slopes with overpotential as well as the reliability of ``equivalent fugacity" relationships from fitting electrochemical permeation data. Its results open up a new avenue for improving the electrochemical techniques generally employed by the hydrogen embrittlement community, bringing the experience gained in electrocatalysis research towards a better understanding and prediction of hydrogen absorption kinetics in cathodically polarized metals susceptible to embrittlement.

\section*{Acknowledgements}
The authors acknowledge financial support from the EPSRC (grant EP/V009680/1) and the EPSRC Henry Royce Institute Materials Challenge Accelerator Programme (grant MCAP019). E. Mart\'{\i}nez-Pa\~neda was additionally supported by an UKRI Future Leaders Fellowship [grant MR/V024124/1].

\section*{Data availability statement}

The data generated during this study will be made available upon reasonable request.


\appendix
\setcounter{table}{0}

\section{Electrochemical permeation data}
\label{Sec:EPdata}

For completeness, the data obtained from the analysis of the permeation transients of this work is provided below for the three electrolytes tested. The data is shown in tables \ref{EPFe} and \ref{EPNi} for, respectively, the iron and nickel membranes.

\begin{table}[H]
 \centering
  \caption{Permeation transients results for Fe. The average diffusion coefficient is calculated disregarding the values below 1$\times10^{-9}$ m$^2$/s and provided along with its 90\% confidence interval.} 
  \label{EPFe}
\resizebox{0.9\textwidth}{!}{
\tiny
\renewcommand{\arraystretch}{0.8}
\begin{tabular}[t]{cccccccc}
\toprule

\text{pH} & \text{Sample} &\text{Transient} & \text{$E$(V$_\text{{Ag/AgCl}})$} & \text{$j_{c}$(mA/cm$^2$)} &  \text{$j_{\infty}$($\upmu$A/cm$^2$)} & \text{$D$(m$^2$/\text{s})}& \text{$C_{0}$(mol/m$^3$)}\\
\midrule
\text{1.3} & \text{a} &\text{1$^{st}$} & \text{-0.54} & \text{0.20} & \text{0.47} & \text{6.7$\times10^{-10}$} & \text{0.024}  \\
\text{1.3} & \text{a} &\text{2$^{nd}$} & \text{-0.58} & \text{0.38} & \text{0.62} & \text{1.8$\times10^{-9}$} & \text{0.031}   \\
\text{1.3} & \text{a} & \text{3$^{rd}$} & \text{-0.62} & \text{0.60} & \text{0.78} & \text{1.9$\times10^{-9}$} & \text{0.039}  \\
\text{1.3} & \text{a} & \text{4$^{th}$} & \text{-0.65} & \text{0.77} & \text{0.90} & \text{2.2$\times10^{-9}$} & \text{0.045}  \\
\text{1.3} & \text{a} &\text{5$^{th}$} & \text{-0.69} & \text{1.00} & \text{1.04} & \text{2.3$\times10^{-9}$} & \text{0.052}  \\
\text{1.3} & \text{a} & \text{6$^{th}$} & \text{-0.74} & \text{1.39} & \text{1.25} & \text{1.9$\times10^{-9}$} & \text{0.063}  \\
\text{1.3} & \text{b} & \text{1$^{st}$} & \text{-0.55} & \text{0.29} & \text{1.20} & \text{1.3$\times10^{-9}$} & \text{0.061}  \\
\text{1.3} & \text{b} & \text{2$^{nd}$} & \text{-0.60} & \text{0.48} & \text{1.41} & \text{2.4$\times10^{-9}$} & \text{0.071}  \\
\text{1.3} & \text{b} & \text{3$^{rd}$} & \text{-0.65} & \text{0.66} & \text{1.70} & \text{2.7$\times10^{-9}$} & \text{0.086}  \\
\text{1.3} & \text{b} & \text{4$^{th}$} & \text{-0.70} & \text{0.87} & \text{2.02} & \text{2.9$\times10^{-9}$} & \text{0.102}  \\
\text{1.3} & \text{b} & \text{5$^{th}$} & \text{-0.75} & \text{1.14} & \text{2.22} & \text{3.4$\times10^{-9}$} & \text{0.112}  \\
\text{1.3} & \text{b} & \text{6$^{th}$} & \text{-0.80} & \text{1.40} & \text{2.44} & \text{3.5$\times10^{-9}$} & \text{0.123}  \\
\text{1.3} & \text{b} & \text{7$^{th}$} & \text{-0.85} & \text{1.66} & \text{2.62} & \text{3.9$\times10^{-9}$} & \text{0.132}  \\
\midrule
\text{7.8} & \text{a} & \text{1$^{st}$} & \text{-0.90} & \text{0.04} & \text{0.49} & \text{1.7$\times10^{-10}$} & \text{0.023}  \\
\text{7.8} & \text{a} &\text{2$^{nd}$} & \text{-1.10} & \text{0.30} & \text{0.76} & \text{3.0$\times10^{-10}$} & \text{0.035}   \\
\text{7.8} & \text{a} &\text{3$^{rd}$} & \text{-1.25} & \text{1.24} & \text{1.32} & \text{1.1$\times10^{-9}$} & \text{0.061}  \\
\text{7.8} & \text{a} & \text{4$^{th}$} & \text{-1.35} & \text{2.29} & \text{1.62} & \text{1.8$\times10^{-9}$} & \text{0.075}  \\
\text{7.8} & \text{a} & \text{5$^{th}$} & \text{-1.45} & \text{3.33} & \text{1.86} & \text{1.9$\times10^{-9}$} & \text{0.086}  \\
\text{7.8} & \text{a} & \text{6$^{th}$} & \text{-1.55} & \text{4.72} & \text{2.04} & \text{2.1$\times10^{-9}$} & \text{0.094}  \\
\text{7.8} & \text{a} & \text{7$^{th}$} & \text{-1.65} & \text{5.92} & \text{2.19} & \text{1.7$\times10^{-9}$} & \text{0.101}  \\
\text{7.8} & \text{b} & \text{1$^{st}$} & \text{-0.90} & \text{0.04} & \text{0.39} & \text{2.4$\times10^{-10}$} & \text{0.018}  \\
\text{7.8} & \text{b} &\text{2$^{nd}$} & \text{-1.20} & \text{0.92} & \text{1.27} & \text{5.5$\times10^{-9}$} & \text{0.059}   \\
\text{7.8} & \text{b} &\text{3$^{rd}$} & \text{-1.40} & \text{2.79} & \text{1.82} & \text{1.7$\times10^{-9}$} & \text{0.084}  \\
\text{7.8} & \text{b} & \text{4$^{th}$} & \text{-1.60} & \text{5.17} & \text{2.11} & \text{2.1$\times10^{-9}$} & \text{0.097}  \\
\midrule
\text{12.8} & \text{a} & \text{1$^{st}$} & \text{-1.20} & \text{0.26} & \text{1.02} & \text{2.4$\times10^{-10}$} & \text{0.052}  \\
\text{12.8} & \text{a} &\text{2$^{nd}$} & \text{-1.30} & \text{0.72} & \text{1.37} & \text{1.0$\times10^{-9}$} & \text{0.069}   \\
\text{12.8} & \text{a} & \text{3$^{rd}$} & \text{-1.40} & \text{1.19} & \text{1.57} & \text{1.1$\times10^{-9}$} & \text{0.080}  \\
\text{12.8} & \text{a} & \text{4$^{th}$} & \text{-1.50} & \text{1.75} & \text{1.70} & \text{1.6$\times10^{-9}$} & \text{0.087}  \\
\text{12.8} & \text{a} & \text{5$^{th}$} & \text{-1.60} & \text{2.34} & \text{1.79} & \text{2.2$\times10^{-9}$} & \text{0.091}  \\
\text{12.8} & \text{a} & \text{6$^{th}$} & \text{-1.80} & \text{3.48} & \text{1.88} & \text{1.9$\times10^{-9}$} & \text{0.096}  \\
\text{12.8} & \text{b} & \text{1$^{st}$} & \text{-1.10} & \text{0.02} & \text{0.47} & \text{3.7$\times10^{-10}$} & \text{0.024}  \\
\text{12.8} & \text{b} &\text{2$^{nd}$} & \text{-1.20} & \text{0.28} & \text{0.95} & \text{4.4$\times10^{-10}$} & \text{0.048}   \\
\text{12.8} & \text{b} & \text{3$^{rd}$} & \text{-1.30} & \text{0.67} & \text{1.34} & \text{1.1$\times10^{-9}$} & \text{0.068}  \\
\text{12.8} & \text{b} & \text{4$^{th}$} & \text{-1.40} & \text{1.14} & \text{1.57} & \text{1.4$\times10^{-9}$} & \text{0.080}  \\
\text{12.8} & \text{b} & \text{5$^{th}$} & \text{-1.50} & \text{1.74} & \text{1.73} & \text{1.4$\times10^{-9}$} & \text{0.088}  \\
\text{12.8} & \text{b} & \text{6$^{th}$} & \text{-1.60} & \text{2.29} & \text{1.84} & \text{2.0$\times10^{-9}$} & \text{0.093}  \\
\midrule
& & & & \text{Average} & \text{2.0$\pm$0.2 $\times 10^{-9}$} &  \\
\bottomrule
\end{tabular}}
\end{table}

\begin{table}[H]
 \centering
  \caption{Permeation transients results for Ni. The average diffusion coefficient is provided along with its 90\% confidence interval.}
  \label{EPNi}
\resizebox{0.9\textwidth}{!}{
\tiny
\renewcommand{\arraystretch}{0.8}
\begin{tabular}[t]{cccccccc}
\toprule

\text{pH} & \text{Sample} & \text{Transient} &\text{$E$(V$_\text{{Ag/AgCl}})$} & \text{$j_{c}$(mA/cm$^2$)} &  \text{$j_{\infty}$($\upmu$A/cm$^2$)} & \text{\it{D}(m$^2$/s)}& \text{$C_{0}$(mol/m$^3$)}\\
\midrule
\text{1.3} & \text{a} & \text{1$^{st}$} & \text{-1.00} & \text{2.50} & \text{0.032} & \text{6.8$\times10^{-14}$} & \text{7.89}  \\
\text{1.3} & \text{a} & \text{2$^{nd}$}  & \text{-1.20} & \text{5.10} & \text{0.069} & \text{1.1$\times10^{-13}$} & \text{16.7}   \\
\text{1.3} & \text{b} & \text{1$^{st}$} & \text{-0.90} & \text{2.45} & \text{0.024} & \text{3.2$\times10^{-14}$} & \text{5.86}  \\
\text{1.3} & \text{b} & \text{2$^{nd}$} &\text{-1.60} & \text{7.51} & \text{0.092} & \text{6.5$\times10^{-14}$} & \text{22.5}  \\
\midrule
\text{7.8} & \text{a} & \text{1$^{st}$} & \text{-1.10} & \text{0.49} & \text{0.090} & \text{1.1$\times10^{-13}$} & \text{22.5}  \\
\text{7.8} & \text{a} & \text{2$^{nd}$} & \text{-1.50} & \text{4.95} & \text{0.156} & \text{1.5$\times10^{-13}$} & \text{39.8}   \\
\text{7.8} & \text{b} & \text{1$^{st}$} & \text{-1.00} & \text{0.52} & \text{0.095} & \text{1.5$\times10^{-13}$} & \text{23.2}  \\
\text{7.8} & \text{b} & \text{2$^{nd}$} & \text{-1.20} & \text{1.10} & \text{0.227} & \text{4.2$\times10^{-13}$} & \text{55.4}  \\
\midrule
\text{12.8} & \text{a} & \text{1$^{st}$} & \text{-1.70} & \text{2.45} & \text{0.177} & \text{3.2$\times10^{-14}$} & \text{44.2}  \\
\text{12.8} & \text{b} & \text{2$^{nd}$} & \text{-2.10} & \text{4.98} & \text{0.247} & \text{5.4$\times10^{-14}$} & \text{61.7}   \\
\text{12.8} & \text{b} & \text{1$^{st}$} & \text{-1.50} & \text{1.05} & \text{0.075} & \text{1.2$\times10^{-13}$} & \text{19.2}  \\
\text{12.8} & \text{b} & \text{2$^{nd}$} & \text{-1.80} & \text{2.23} & \text{0.285} & \text{1.3$\times10^{-13}$} & \text{72.8}  \\
\midrule
& & & & \text{Average} & \text{9.1$\pm$4.7 $\times 10^{-14}$} &  \\
\bottomrule
\end{tabular}}
\end{table}

\section{Steady-state permeation slopes} \label{app:B}
For completeness, we proceed to show how Eqs. (\ref{eq:jinf3b}) and (\ref{eq:jinf3c}) follow from the species balance of surface adsorbed hydrogen ($\theta_\text{ads}$), and the imposed electric current and charge conservation. The two conditions that need to be fulfilled, in their general form, are prescribed as the conservation of surface-adsorbed hydrogen:
\begin{equation} \label{eq:appB_thetaBalance}
    \frac{\partial \theta_\text{ads}}{\partial t} = \nu_\text{Va} + \nu_\text{Vb} - \nu_\text{Ha} - \nu_\text{Hb} - 2\nu_\text{T} - \frac{j_\infty}{F} = 0
\end{equation}
and the electric current involved within the reactions:
\begin{equation} \label{eq:appB_iBalance}
     0 = \nu_\text{Va} + \nu_\text{Vb} + \nu_\text{Ha} + \nu_\text{Hb} - \frac{j_\text{ext}}{F}
\end{equation}
with the reaction rates from the individual reaction steps given by \cref{eq:vVa,eq:vHa,eq:vVb,eq:vHb,eq:vT}, and the electric current $j_\text{ext}$ being the current required to maintain the electric potential imposed by the experimental setup. We assume that the permeation current is low, such that $j_\infty/F<<\nu$, allowing the permeation current to be neglected relative to other reaction rates. Finally, we note that based on Eq. (\ref{eq:jinf2}), the following holds:
\begin{equation} \label{eq:appB_derivj}
    \frac{\partial \text{ln}(j_\infty)}{\partial \eta} = \frac{\partial \text{ln}\left(k_\text{A}\theta_\text{ads}-k_\text{A}'C(\theta_\text{ads})\right)}{\partial \theta_\text{ads}}\frac{\partial \theta_\text{ads}}{\partial \eta} = \frac{1}{\theta}\frac{\partial \theta_\text{ads}}{\partial \eta}  = \frac{\partial \text{ln}(\theta_\text{ads})}{\partial \eta}
\end{equation}
resulting in Eq. (\ref{eq:jinf3}). As a result, the slope of $\text{ln}(\theta_\text{ads})$ only depends on the surface coverage and its scaling with the overpotential, and is independent of the surface adsorption process. 
\subsection{Volmer-Heyrovsky}
When the Heyrovsky reaction step is the dominant one, the balance of surface-adsorbed hydrogen, Eq. (\ref{eq:appB_thetaBalance}), reduces to:
\begin{equation}
    0 = \left(k_\text{Va}C_{\text{H}^+}+k_\text{Vb}\right)(1-\theta_\text{ads})\text{exp}\left(\frac{-\beta_\text{V}\eta F}{RT}\right) - \left(k_\text{Ha}C_{\text{H}^+}+k_\text{Hb}\right)\theta_\text{ads} \text{exp}\left(\frac{-\beta_\text{H}\eta F}{RT}\right)
\end{equation}
where it is assumed that the acidic and basic reaction steps use the same symmetry coefficient, $\beta_\text{Va}=\beta_\text{Vb}=\beta_\text{V}$ and $\beta_\text{Ha}=\beta_\text{Hb}=\beta_\text{H}$. For small surface coverage, $\theta_\text{ads}<<1$, it follows that:
\begin{equation}
    \theta_\text{ads} = \text{exp}\left(\frac{(\beta_\text{H}-\beta_\text{V})\eta F}{RT}\right) \frac{k_\text{Va}C_{\text{H}^+}+k_\text{Vb}}{k_\text{Ha}C_{\text{H}^+}+k_\text{Hb}}
\end{equation}
From which it then follows through Eq. (\ref{eq:appB_derivj}) that:
\begin{equation}
    \frac{\partial \text{ln}(j_\infty)}{\partial \eta} = \frac{1}{\theta} \frac{\partial \theta}{\partial \eta} = \frac{(\beta_\text{H}-\beta_\text{V})F}{RT}
\end{equation}
Resulting in the expression from Eq. (\ref{eq:jinf3b}), providing a constant slope indicative of the difference between symmetry coefficients. For completeness, the current conservation can be used to provide the electric current supplied by the imposed overpotential via Eq. (\ref{eq:appB_iBalance}) as:
\begin{equation}
    j_\text{ext} = 2 F (1-\theta_\text{ads})\left(k_\text{Va}C_{\text{H}^+}+k_\text{Vb}\right) \text{exp}\left(\frac{-\beta_\text{V}\eta F}{RT}\right) = 2F\left(k_\text{Ha}C_{\text{H}^+}+k_\text{Hb}\right)\theta_\text{ads} \text{exp}\left(\frac{-\beta_\text{H}\eta F}{RT}\right)
\end{equation}
where equal reaction rates between the Volmer and Heyrovsky reactions indicate that the total electric current used to sustain the reactions is twice that of one of the individual reaction currents. It should be noted that, under lab conditions, this current is directly supplied by the voltage supply that is included within the experimental setup to impose the electric overpotential, allowing this potential to deviate from its equilibrium potential. For practical applications, this current would also be supplied externally (e.g. in the case of hydrogen gas production) or the result of anodic corrosion reactions occurring on the metal surface (or on connected metal surfaces). 
\subsection{Volmer-Tafel}
If instead the Tafel reaction step is the dominant desorption mechanism for surface-adsorbed hydrogen, the surface balance from Eq. (\ref{eq:appB_thetaBalance}) is given as:
\begin{equation}
    0=\left(k_\text{Va}C_{\text{H}^+}+k_\text{Vb}\right)(1-\theta_\text{ads})\text{exp}\left(\frac{-\beta_\text{V}\eta F}{RT}\right) - 2 k_\text{T} \theta_\text{ads}^2
\end{equation}
such that the surface coverage is given by:
\begin{equation}
    \theta_\text{ads} = \left(\frac{k_\text{Va}C_{\text{H}^+}+k_\text{Vb}}{2 k_\text{T}}\right)^{\frac{1}{2}} \text{exp}\left(\frac{-\beta_\text{V}\eta F}{2RT}\right)
\end{equation}
and the hydrogen permeation current reads:
\begin{equation}
    \frac{\partial \text{ln}(j_\infty)}{\partial \eta} = \frac{1}{\theta} \frac{\partial \theta}{\partial \eta} = \frac{-\beta_\text{V}F}{2RT}
\end{equation}
providing Eq. (\ref{eq:jinf3c}). For completeness, the electric current imposed to sustain the required electric overpotential is given through Eq. (\ref{eq:appB_iBalance}) as:
\begin{equation}
    j_\text{ext} = F (1-\theta_\text{ads})\left(k_\text{Va}C_{\text{H}^+}+k_\text{Vb}\right) \text{exp}\left(\frac{-\beta_\text{V}\eta F}{RT}\right)
\end{equation}


\end{document}